\documentclass[prd,showpacs,showkeys,preprintnumbers,floatfix,groupedaddress,
nofootinbib,superscriptaddress]{revtex4-1}
\usepackage{hyperref} 
\usepackage{graphicx}
\usepackage{amssymb}
\usepackage{amsmath}
\usepackage{bbold}
\usepackage{epstopdf}
\usepackage{float}
\usepackage{caption}
\usepackage{slashed}
\usepackage{verbatim}
\usepackage[singlelinecheck=false]{subcaption}

\DeclareGraphicsExtensions{.pdf,.png,.jpg}
\DeclareMathOperator{\tr}{tr}
\def\chpt{\raise0.4ex\hbox{$\chi$}PT}
\def\mhat{\widehat m}
\def\muhat{\widehat \mu}

\begin{document}

\title{Phase diagram of non-degenerate twisted mass fermions}

\author{Derek P. Horkel}
\email[e-mail: ]{dhorkel@uw.edu}
\affiliation{
 Physics Department, University of Washington, 
 Seattle, WA 98195-1560, USA \\
}
\author{Stephen R. Sharpe}
\email[e-mail: ]{srsharpe@uw.edu}
\affiliation{
 Physics Department, University of Washington, 
 Seattle, WA 98195-1560, USA \\
}
\date{\today}
\begin{abstract}
We determine the phase diagram and pion spectrum for
Wilson and twisted-mass fermions in the presence of non-degeneracy
between the up and down quark {\em and} discretization errors,
using Wilson and twisted-mass chiral perturbation theory.
We find that the CP-violating phase of the continuum theory
(which occurs for sufficiently large non-degeneracy)
is continuously connected to the Aoki phase of the
lattice theory with degenerate quarks.
We show that discretization effects can, in some cases, push simulations
with physical masses closer to either the CP-violating phase
or another phase not present in the continuum, so that
at sufficiently large lattice spacings physical-point simulations
could lie in one of these phases.
\end{abstract}

\maketitle

\section{Introduction}
\label{sec:intro}

It has long been known, in the case of three light quarks,
that there is a transition to a CP-violating phase for 
non-degenerate quarks when
one of the quark masses becomes sufficiently negative~\cite{Dashen:1970et}.
For example, using leading order (LO)
SU(3) chiral perturbation theory (\chpt), and fixing $m_d$ and $m_s$,
the transition occurs when $m_u=-m_d m_s/(m_d+m_s)$~\cite{Creutz:2003xu}.
The neutral pion becomes massless on the transition line,
and within the new phase the chiral order parameter, $\langle\Sigma\rangle$,
becomes complex.
For physical QCD this is mostly a curiosity, since increasingly
accurate determinations of the quark masses indicate clearly that
all are positive relative to one 
another~\cite{Beringer:1900zz,Aoki:2013ldr}.
Thus physical QCD, despite the non-degeneracy of
the up and down quarks, lies away from the critical line.

For lattice QCD (LQCD), however, the situation is less clear.
The position of the transition can be shifted 
closer to the physical point by discretization effects. 
Indeed, it is well known that,
with degenerate Wilson-like fermions,\footnote{%
``Wilson-like'' indicates that the analysis holds for both Wilson fermions
and various improvements thereof, in particular for non-perturbatively
${\cal O}(a)$-improved Wilson fermions.}
discretization effects can lead
to the appearance of a new phase---the Aoki phase---in which
isospin is spontaneously broken
and $\langle\Sigma\rangle$ is complex~\cite{Aoki:1983qi,Sharpe:1998xm}.
In addition, advances in simulations now allow calculations to be done at 
the physical light-quark masses, including, very recently,
the physical non-degeneracy between up and down quarks~\cite{Borsanyi:2014jba}.
It is thus natural to ask how, in LQCD with
non-degenerate quarks, discretization effects change
the position and nature of the CP-violating phase.
This question is particularly acute in the case of twisted-mass
fermions, where additional symmetry breaking is explicitly included.

In this paper we address this question for Wilson-like
and twisted-mass lattice fermions.
We do so using \chpt,
specifically the versions of \chpt\
in which the effects of discretization have been included.
Our work also allows us to address a related issue:
In what way is the CP-violating phase of the continuum theory related to
the Aoki phase of the lattice theory?\footnote{%
This issue has been raised previously by Mike Creutz
and his conjectured answer is confirmed by the present 
analysis~\cite{Creutz:2014em}.}

Since twisted-mass QCD is only defined for even numbers of fermion
flavors~\cite{Frezzotti:2003ni}, a necessary step for our work
is to rephrase the continuum SU(3) \chpt\ analysis of 
Ref.~\cite{Creutz:2003xu} in the two-flavor theory 
obtained by integrating out the strange quark.
This requires that the contributions
of one of the next-to-leading order (NLO)
low-energy coefficients ($\ell_7$) be treated as parametrically
larger than the others.
Thus we are led to a somewhat non-standard power-counting,
but one which reproduces the SU(3) phase diagram, including the
CP-violating phase, within SU(2) \chpt.
This approach has been used before along the line 
$m_u=-m_d$~\cite{Smilga:1998dh};
here we extend the analysis to arbitrary mass splitting. 
Similar work has also been done recently in the context
of a effective theory including the $\eta$ meson~\cite{Aoki:2014moa}.

The organization of this article is as follows.
In Sec.~\ref{sec:vacuum} we briefly recall the results for the
phase structure and pion masses at LO in SU(2) and SU(3) \chpt,
and show how they differ.
Section~\ref{sec:matching}
describes the matching of SU(3) and SU(2) \chpt.
In Sec.~\ref{sec:disc}, we recall briefly how
discretization effects are incorporated in \chpt\
for degenerate Wilson-like fermions, and the resulting phase structure.
We then present our first new results:
the phase diagram including both discretization effects and non-degeneracy.
In Sec.~\ref{sec:twist} we move onto twisted-mass fermions,
focusing first on the phase diagram and pion masses in the
case of maximal twist, where most simulations have been done 
because of the property of automatic
$\mathcal{O}(a)$ improvement~\cite{Frezzotti:2003ni}.
It is nevertheless interesting to understand how the results 
with untwisted and maximally twisted fermions are connected,
and so, in Sec.~\ref{sec:arbtwist}, we discuss the phase diagram for
general twist.

Up to this stage, our analysis is done using the LO terms due to
the average quark mass, discretization effects and non-degenerate quark masses.
To understand how robust the results are we consider,
in Sec.~\ref{sec:higher}, the impact of including the next higher order
terms in our power counting.
Some conclusions are collected in Section \ref{sec:concl}.

\section{Continuum Vacuum Structure at leading order in \chpt}
\label{sec:vacuum}

In this section we review the vacuum structure predicted by LO
\chpt\ for both two and three light flavors.
The LO chiral Lagrangian in Euclidean space-time is, for any
number of light flavors,
\begin{equation}
\mathcal{L}_\chi = \frac{f^2}{4}\tr\left[
\partial_\mu \Sigma \partial_\mu \Sigma^\dagger
-(\chi\Sigma^\dagger+\Sigma\chi^\dagger)\right]\,,
\end{equation}
where $\Sigma\in SU(N_f)$ and $\chi=2B_0 M$
(with $M$ the mass matrix), while
$f\sim 92\;$MeV and $B_0$ are low-energy constants (LECs).

\begin{figure}[tb!]
\centering
\includegraphics[scale=.3]{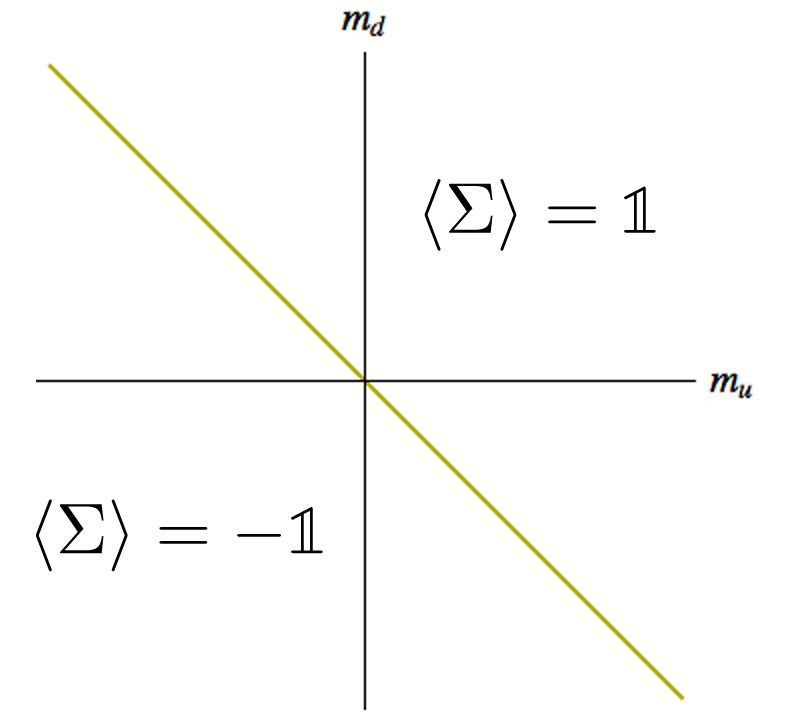}
\caption{\label{fig:SU2LO} Phase diagram at lowest order in SU(2) \chpt.}
\end{figure}

For two light flavors the chiral order parameter can be parametrized
as $\langle\Sigma\rangle=\exp(i\theta \hat n\cdot \vec\tau)$. Although the
mass matrix $M={\rm diag}(m_u,m_d)$ has both singlet and triplet components,
the leading order potential depends only on the former
\begin{equation}
\mathcal{V}_{SU(2),\,LO} 
= -\frac{f^2}{4}\tr\left[\chi\Sigma^\dagger+\Sigma\chi^\dagger\right]
= -\frac{f^2}{2}\cos{\theta}\tr[\chi] 
\equiv -f^2 \cos{\theta}\,\chi_\ell\,.
\end{equation}
In the last step we have defined the convenient quantity 
$\chi_\ell=B_0 (m_u+m_d)$.
The potential is minimized at $\theta=0$ if $\chi_\ell >0$ 
and at $\theta=\pi$ if $\chi_\ell<0$, resulting in
the phase diagram sketched in Fig.~\ref{fig:SU2LO}. 
In terms of the behavior of the condensate,
this is a first-order phase transition at
which the condensate flips sign.
This characterization is somewhat misleading, however,
because the two sides of the transition are related by
a non-anomalous flavor rotation. Such a transformation
can change $M \to -M$ and $\Sigma\to-\Sigma$, 
while leaving physics unchanged. Thus by adding an extra
dimension to the phase diagram (as we will do later) one
finds that the two sides are connected.

Expanding the potential about its minimum,
using $\Sigma=\langle \Sigma\rangle\exp(i\vec \pi\cdot \vec \tau/f)$
we find the standard LO result for the pion masses, 
$m^2_\pi=|\chi_\ell|$. 
These thus vanish along the phase transition line.
That they vanish at the origin follows from Goldstone's theorem due
to the spontaneous breaking of the exact axial symmetry. 
That they vanish away from
the origin along the transition line is not expected from
symmetry arguments, and indeed holds, as we will see, only at LO in \chpt.

\begin{figure}[bt!]
\centering
\includegraphics[scale=.3]{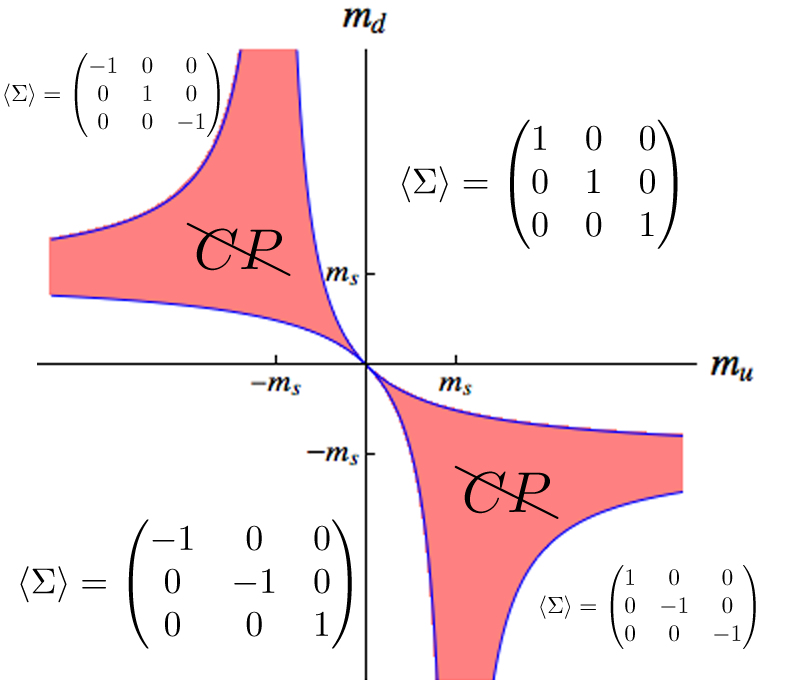}
\caption{\label{fig:SU3LO} Phase diagram at lowest order in SU(3) \chpt\ 
with fixed strange quark mass. 
Equations for the positions of phase transition lines are given in the text.} 
\end{figure}

The phase diagram of the three-flavor theory has a more interesting
structure, as elucidated most extensively by Creutz~\cite{Creutz:2003xu}.
Since $m_s\gg m_u,m_d$ in nature, it is natural to hold $m_s$ fixed and
vary the other two quark masses.
The resulting phase diagram at LO is sketched in Fig.~\ref{fig:SU3LO}.
The ``normal'' region, in which $\langle\Sigma\rangle=\mathbb 1$, ends
at a transition line along which $m_{\pi^0}$ vanishes.
This occurs (for fixed $m_s>0$) when one of the other masses,
say $m_u$, becomes sufficiently negative.
The explicit expression for the neutral pion mass in this phase is
 \begin{equation}
m_{\pi^0\, SU(3)}^2=\frac{2}{3}B_0\left(m_u+m_d+m_s 
- \sqrt{m_u^2+m_d^2+m_s^2-m_u m_d-m_u m_s-m_d m_s}\right)\,,
\label{eq:mpi0su3}
\end{equation}
which vanishes when $m_u=-m_d m_s/(m_d + m_s)$.
The charged pions remain massive throughout the normal phase except
at the origin.

Moving outside the normal phase one enters a CP-violating phase in which
the condensate is complex. The explicit form is
\begin{equation}
\left<\Sigma\right>=\begin{pmatrix}
\exp{i\phi} & 0 & 0 \\
0 & \exp{i\psi} & 0 \\
0 & 0 & \exp{-i(\phi+\psi)}
\end{pmatrix}
\end{equation}
where the phases satisfy
\begin{equation}
m_u\sin{\phi}=m_d\sin{\psi}=-m_s\sin{(\phi+\psi)}
\,.
\end{equation}
In this case there is a genuine phase transition at the boundary.
It is of second order: 
$\langle\Sigma\rangle$ is continuous, and a single pion becomes massless.

The phase diagram is symmetric under both
$m_u\leftrightarrow m_d$ interchange
and inversion through the origin (with $m_s$ fixed).
Inversion is brought about by a non-anomalous axial isospin
transformation, which also changes the condensate as shown 
in Fig.~\ref{fig:SU3LO}.
We note that the CP-violating region is of finite width.\footnote{%
The theory along the $m_u=-m_d$ diagonal is identical to that
with $m_u=m_d$ at $\theta_{\rm QCD}=\pi$, and has been discussed
extensively in the literature. In particular, a \chpt\ analysis
of this theory has been given in Ref.~\cite{Smilga:1998dh}.}
Specifically,
as one moves away from the origin along the $m_u=-m_d$ diagonal,
the width of this region grows proportionally to $(m_u\!-\!m_d)^2/m_s$.

As the figure shows, there are additional phase boundaries
in the second and fourth quadrants. These occur, however, when
$|m_u|,|m_d|> |m_s|$, and thus lie far from the region of physical interest.
In the rest of our analysis, we consider only the region in which
$|m_u|,|m_d| \ll |m_s|$, and thus zoom in on the vicinity of the origin
in Fig.~\ref{fig:SU3LO}. 

\section{Matching SU(2) and SU(3) \chpt\  for non-degenerate quarks}
\label{sec:matching}

If we choose the quark masses to satisfy
$|m_u|,|m_d| \ll |m_s| \ll \Lambda_{\rm QCD}$,
then the properties of pions can be simultaneously described
by both SU(2) and SU(3) \chpt,
and the predictions of the two theories must agree. 
The results of the previous section show that this is 
not the case if we work to LO in both theories---the 
CP-violating phase is absent in SU(2) \chpt. 
The discrepancy is resolved by noting that
the CP-violating phase has a width proportional to 
$(m_u\!-\!m_d)^2$, indicating that it arises at NLO in SU(2) \chpt. 
In this section we recall how the
two theories are matched, and show how the CP-violating phase can
then be obtained in SU(2) \chpt\ when including the resulting NLO term.

To do the matching, one considers quantities 
accessible in both SU(2) and SU(3) theories,
namely pion masses and scattering amplitudes.
Expanding the LO SU(3) result in powers of
$m_{u,d}/m_s$, the leading terms match with the LO SU(2) result,
while the first subleading terms match with an NLO SU(2) contribution.
The subleading terms in the SU(3) results
are in fact proportional to $(m_u\!-\!m_d)^2$,
because they arise from intermediate $\eta$ propagators
and involve two factors of the $\pi^0-\eta$ mixing amplitude.
The only source of such mass dependence at NLO in the SU(2) theory
is the $\ell_7$ term in the NLO potential
\begin{equation}
\mathcal{V}_{SU(2)\, NLO} =
-\frac{\ell_3}{16} [\tr(\chi^\dagger \Sigma + \Sigma^\dagger \chi)]^2 
+\frac{\ell_7}{16}[\tr(\chi^\dagger \Sigma - \Sigma^\dagger \chi)]^2\,.
\end{equation}
Writing $\chi$ as
\begin{equation}
\chi = \chi_\ell \mathbb 1 + \epsilon \tau_3\,, \ \ {\rm with}\ \ 
\epsilon= B_0(m_u-m_d)\,,
\end{equation}
we see that only the $\epsilon$ part contributes to the $\ell_7$
term. Thus this term leads to contributions proportional to
$(m_u\!-\!m_d)^2$.
Other NLO contributions (i.e. those proportional to different
NLO LECs or coming from loops) do not have this mass dependence.

The simplest quantity with which to do the matching is
the neutral pion mass, and this was used to determine the
value of $\ell_7$ in Ref.~\cite{Gasser:1984gg}.
The LO SU(3) result [given in Eq.~(\ref{eq:mpi0su3}) above] expands to
\begin{equation}
m_{\pi^0 SU(3)\, LO}^2=\chi_\ell -\frac{\epsilon^2}{4B_0 m_s} 
+\mathcal{O}\left(\frac{\epsilon^2 m_{u,d}}{m_s^2}\right)\,.
\end{equation}
The SU(2) result at NLO is
\begin{equation}
m_{\pi^0 SU(2)\, NLO}^2=\chi_\ell - \frac{2\ell_7 \epsilon^2 }{f^2}
+ {\cal O}\left(\frac{\chi_\ell^2}{\Lambda_\chi^2}\right)\,,
\end{equation}
where $\Lambda_\chi=4\pi f$ is the chiral scale.
The $\chi_\ell^2$ contributions arise from terms in the NLO
chiral Lagrangian (including $\ell_3$) as well as from chiral logarithms.
Equating these two results one finds~\cite{Gasser:1984gg}
\begin{equation}
\ell_7=\frac{f^2}{8B_0m_s}\,.
\label{eq:ell7match}
\end{equation}
One can show that with this value for $\ell_7$,
contributions to all pion $n$-point amplitudes 
proportional to $\epsilon^2/m_s$ agree in the two theories.

We stress that in this matching we are not taking
into account ``standard'' NLO contributions, i.e. those
suppressed relative to LO results by factors of
$m_{u,d}/\Lambda_{\rm QCD}\sim (m_\pi/\Lambda_\chi)^2$
(up to logarithms). Such contributions arise in both
SU(3) and SU(2) \chpt\ and must be included in a full NLO matching.
This is not necessary for our purposes since such terms lead
to small isospin-conserving corrections to the vacuum structure
and pion masses---they do not introduce qualitatively new effects.
By contrast, the $\epsilon^2$ terms that we keep lead to
isospin breaking, and are the leading order contributions which do so.
Indeed, for this reason $\ell_7$ is not renormalized at this order,
since, as already noted, one-loop 
chiral logarithms do not contain a term proportional to $\epsilon^2$.
Thus it is consistent to work with the classical potential,
rather than the one-loop effective potential.
This is not the case for other LECs such as $\ell_3$,
which are renormalized and thus scale-dependent~\cite{Gasser:1984gg}.

We can formalize this by noting that standard NLO contributions
are parametrically smaller than the terms we keep by a factor of 
$m_s/\Lambda_{\rm QCD}$. This allows the development of a consistent
power-counting scheme in which the $\epsilon^2$ terms are larger
than generic $m^2$ contributions.\footnote{%
The numerical basis for this power-counting is not very strong.
For example, $\ell_7$ and $\ell_3(\mu)$ are comparable in size
for reasonable values of the scale $\mu$.
Thus the numerical size of the standard NLO corrections we
are dropping may be comparable to those proportional to $\epsilon^2$
that we are keeping.
The key point, however, is that we are interested in qualitatively
new effects, rather than a precise quantitative description.}
We discuss this in the following
section. To be consistent we should also account for NLO
contributions in SU(3) \chpt\ of size $m_s/\Lambda_{\rm QCD}$ relative 
to LO terms. These, however, lead only to a renormalization of the
SU(2) constants $f$ and $B_0$ relative to their SU(3) counterparts.
Since we work henceforth entirely in the SU(2) theory,
we choose to leave this renormalization implicit.

We now show that the inclusion of the $\ell_7$ term leads to the
same phase diagram as found in the LO SU(3) analysis.
Given the matching result Eq.~(\ref{eq:ell7match}), we 
always assume $\ell_7>0$ in the following.
Using $\langle\Sigma\rangle=\exp(i\theta \hat n\cdot \vec \tau)$, 
the potential becomes
\begin{equation}
\mathcal{V}_{SU(2)} = - f^2 \left( \chi_\ell \cos{\theta} 
+c_\ell \epsilon^2 n_3^2 \sin^2{\theta}\right)\,,
\label{eq:V2NLO}
\end{equation}
where $c_\ell=\ell_7/f^2$.
Since $\ell_7>0$, the potential is always minimized by choosing
$|n_3|=1$. Since $n_3=1$ and $n_3=-1$ are related by changing
the sign of $\theta$, we can, without loss of generality, set $n_3=1$.
The resulting potential is stationary with respect to $\theta$
at the ``normal'' values $\theta=0$ and $\pi$,
and in addition at
\begin{equation}
\cos\theta = \frac{\chi_\ell }{2c_\ell\epsilon^2}\,.
\end{equation}
This new stationary value always leads to the global minimum
of the potential where it is valid,
i.e. when $|\cos\theta|\le 1$.
Thus, for fixed $\epsilon$, there is a new
phase for $-2c_\ell\epsilon^2 \le \chi_\ell \le 2 c_\ell\epsilon^2$,
within which $\langle\Sigma\rangle$ is complex and CP is violated.
Although $\cos\theta$ is fixed, the sign of $\theta$ is not, with the
two possible vacua begin related by a CP transformation.
This phase matches continuously onto the normal phases with $\cos\theta=\pm1$
at its boundaries. Thus the phase transition is of second order.

\begin{figure}[tb!]
\centering
\includegraphics[scale=.3]{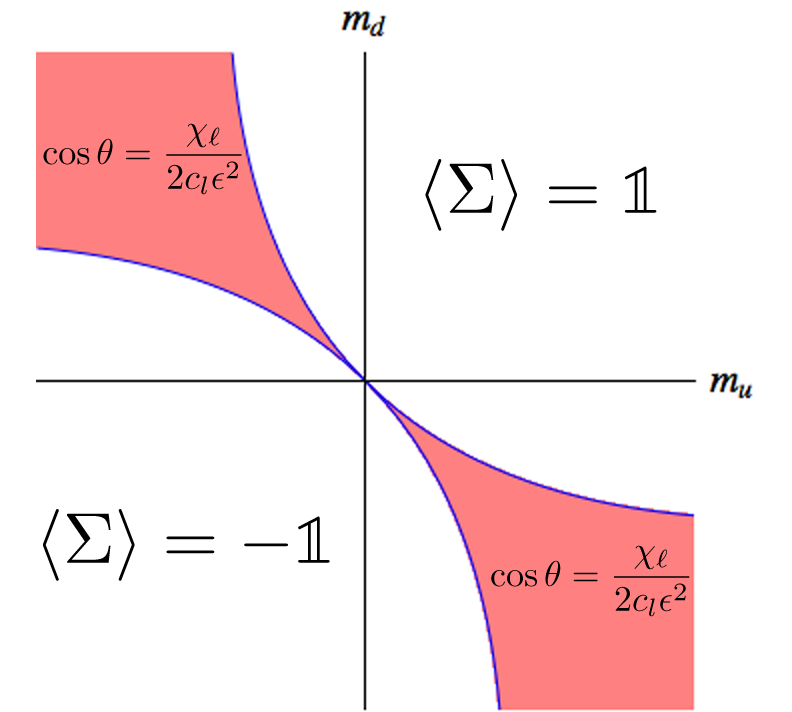}
\caption{\label{fig:NLOell7} Phase diagram from SU(2) \chpt\ 
including $\ell_7$ term with $\ell_7>0$. 
Equations for the positions of phase transition lines are given in the text.} 
\end{figure}

The resulting phase diagram is sketched in Fig.~\ref{fig:NLOell7}.
This is not only qualitatively similar to the central portion
of the LO SU(3) phase diagram, Fig.~\ref{fig:SU3LO}, 
but is in fact in complete quantitative agreement at the appropriate order.
For example, expanding the SU(3) result for the phase boundary,
$m_u=-m_d/(1+m_d/m_s)$, in powers of $m_{u,d}/m_s$, and keeping only
the leading non-trivial term,  one finds
that the boundary occurs at $\chi_\ell=\epsilon^2/(4 B_0 m_s)$.
This agrees with the SU(2) result $\chi_\ell=2 \ell_7 \epsilon^2/f^2$ 
using the matching condition (\ref{eq:ell7match}).
We have also checked that the pion masses agree throughout the phase plane.
We do not quote results for pion masses here,
since they are included in the more general analysis presented below.

The fact that the CP-violating phase can be reproduced within SU(2) \chpt\
was first explained by Smilga~\cite{Smilga:1998dh}. His work considered
only the case $m_u=-m_d$, which, as noted above, is the same as
$m_u=m_d$ with $\theta_{\rm QCD}=\pi$. The analysis presented here gives
the (very simple) generalization to arbitrary non-degenerate quark masses.
There is also a close relation between our analysis and the recent work
of Aoki and Creutz~\cite{Aoki:2014moa}.
These authors do not use \chpt\ {\em per se}, but rather
an effective theory containing both pions and the $\eta$ meson.
If the $\eta$ were integrated out then their theory would reduce to that
we consider here, including the $\ell_7$ term, plus small corrections.
We think, however, that it is preferable to work in a strict effective theory
framework, in which only the light particles are kept as dynamical degrees
of freedom.

\section{Including discretization effects for Wilson-like fermions}
\label{sec:disc}

In this section we recall how lattice artifacts 
can be incorporated into \chpt, and study their impact on the
phase structure described above at leading non-trivial order.
We do this for untwisted Wilson-like fermions---twist 
will be considered in the following sections.
The method leads to the chiral
effective theory describing lattice simulations close to the continuum limit.
We begin by recalling the analysis for degenerate quarks and then
add in non-degeneracy.
We work entirely in the two-flavor theory obtained after the strange quark 
(and the charm quark too, if present) has been integrated out.
For untwisted Wilson-like fermions (unlike for twisted-mass fermions), 
the analysis could also be carried out within SU(3) \chpt, 
but there is no advantage to doing so as the dominant long-distance
dynamics lies in the SU(2) sector.

Both quark masses and discretization effects break chiral symmetry,
and it is important to understand the relative size of these effects. 
Our focus here is on state-of-the-art simulations, which have $m_{u,d}$ close
to their physical values ($m_u\approx 2.5\;$MeV and $m_d\approx 5\;$MeV
in the $\overline{\rm MS}$ scheme at $\mu=2\;$GeV),
and lattice spacings such that $1/a\approx 3\;$GeV.
In this case, the relative size of discretization effects is characterized
by $a \Lambda_{\rm QCD}\approx 0.1$ (using $\Lambda_{\rm QCD}=300\;$MeV),
so that
\begin{equation}
a \Lambda_{\rm QCD}^2\approx 30\,{\rm MeV} \gg m_{u,d} 
\approx a^2 \Lambda_{\rm QCD}^3 \approx 3\,{\rm MeV}\,.
\end{equation}
The appropriate power-counting is thus (in schematic notation) $a^2 \sim m$.
This is the Aoki regime, in which competition between discretization
and mass effects leads to interesting phase 
structure~\cite{Aoki:1983qi,Sharpe:1998xm}.

Discretization effects can be incorporated into \chpt\
following the method of Ref.~\cite{Sharpe:1998xm}. 
For unimproved (or partially improved) Wilson fermions,
the dominant discretization effect is proportional to $a$.
In the pion sector, however, this contribution
can be absorbed entirely into a common shift
in all quark masses~\cite{Sharpe:1998xm}, and we assume below
that this shift has been made. 
The first non-trivial discretization effect is that proportional to $a^2$.
This changes the LO potential to~\cite{Sharpe:1998xm}
\begin{align}
\mathcal{V}_{a^2}=&-\frac{f^2}{4} 
\tr({\chi}^\dagger \Sigma + \Sigma^\dagger {\chi})
- W' [\tr(\hat A^\dagger \Sigma +   \Sigma^\dagger \hat A)]^2\,.
\end{align}
Here we are using the notation of Ref.~\cite{Sharpe:2004ps},
in which $\hat A=2 W_0 a \mathbb 1$ is a spurion field, 
with dimensions of mass squared,
and proportional to the identity matrix in flavor space.
$W_0$ and $W'$ are new LECs.

\begin{figure}[tb!]
\centering
\includegraphics[scale=.3]{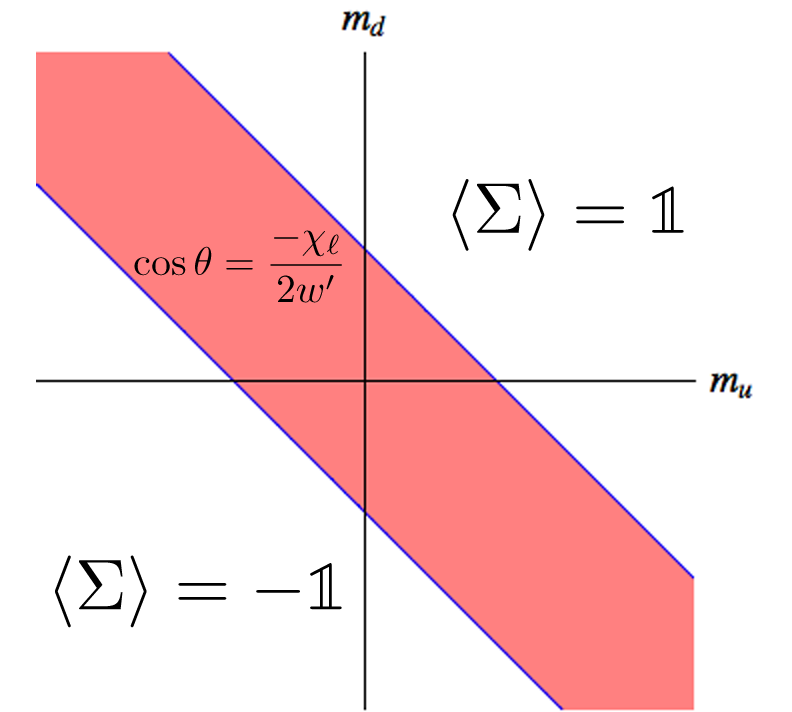}
\caption{\label{fig:AokiLO} Phase diagram in LO SU(2) \chpt\ including
discretization effects with $w'<0$ (Aoki scenario).
Equations for the positions of phase transition lines are given in the text.}
\end{figure}

The analysis of the vacuum structure for degenerate quarks was given in
Ref.~\cite{Sharpe:1998xm}. Since ${\cal V}_{a^2}$ is independent
of the $\epsilon$, the results are
unchanged at LO in the presence of non-degeneracy.
To determine the vacuum we must minimize
\begin{equation}
\mathcal{V}_{a^2} = - f^2 
\left(\chi_\ell \cos{\theta} +w' \cos^2{\theta}\right)\,,
\end{equation} 
where $w'={64W'W_0^2 a^2}/{f^2}$.
For $w'<0$, the analysis is essentially the same as that
for ${\cal V}_{SU(2)}$ with $\ell_7>0$, as given in the previous section.
Stationary points are at $\cos\theta=\pm 1$ and
\begin{equation}
\cos{\theta}=-\frac{\chi_\ell}{2w'}\,,
\end{equation}
with the latter being the global minimum  where valid ($|\cos\theta|\le 1$).
This leads to the phase diagram shown in Fig.~\ref{fig:AokiLO},
with an Aoki phase~\cite{Aoki:1983qi} 
separated from the normal phases by second-order transitions
at $|\chi_\ell|=-2w'$.
Strictly speaking, the name ``Aoki phase'' has been applied previously
only on the diagonal $m_u=m_d$ axis, but in the present approximation
it holds also for non-degenerate quarks.
Within the Aoki phase the potential
is independent of the direction of the condensate, $\hat n$,
so that there are two massless Goldstone bosons, the charged pions.
Parity and flavor are violated within this phase. With the
canonical choice of the direction of the condensate, 
$\hat n=\hat z$, CP is also violated.

For $w'>0$, the global minimum lies at $\cos\theta={\rm sign}( \chi_\ell)$,
with a first-order transition at $\chi_\ell=0$.
The phase diagram is thus identical to that in the continuum,
Fig.~\ref{fig:SU2LO}.
The only difference is that here the yellow line indicates
a genuine first-order transition, since on the lattice there are
no symmetries connecting the two sides.
This case is referred to as the first-order scenario~\cite{Sharpe:1998xm}.

\bigskip
We are now ready to combine the effects of non-degeneracy with
discretization errors. This requires that we adopt an appropriate
power-counting scheme for the relative importance of 
$\epsilon^2$, $m$ and $a^2$, where $m$ indicates a generic quark mass.
Recalling that $\epsilon^2$ terms are enhanced compared to generic $m^2$ terms
we use
\begin{equation}
m\sim a^2 > \epsilon^2 > m a \sim a^3 > a \epsilon^2 > 
m^2\sim m a^2 \sim a^4 \dots.
\end{equation}
This can be thought of as treating $\epsilon\sim a^{1+\delta}$,
with $0<\delta < 1/2$.
The utility of this power counting is that allows us to
first add the $\epsilon^2$ term to those proportional to $m$ and $a^2$,
and then consider terms of order $ma\sim a^3$ at a later stage
(in Sec.~\ref{sec:higher} below).
Indeed, we could, for the purposes of this section, set $\delta=0$,
and treat the $\epsilon^2$ term as of LO.
We do not do so, however, since this would 
require us to later treat $a\epsilon^2$ terms as of the same size
as those proportional to $ma\sim a^3$.
Nevertheless, we will loosely describe the inclusion
of $m$, $a^2$ and $\epsilon^2$ terms as constituting our LO analysis,
while treating the $ma\sim a^3$ terms as being of NLO. Terms of yet higher
order will not be considered.

With the power counting in hand, we can extend the inclusion of
discretization errors into \chpt\ to incorporate the effects of non-degeneracy.
This leads to the appearance of new operators in the Symanzik effective
Lagrangian, and thus, potentially, to new terms in the chiral Lagrangian.
The constraints on additional operators in the Symanzik Lagrangian
in the presence of non-degeneracy 
were worked out in Ref.~\cite{WalkerLoud:2005bt}.
Using their results within our power-counting scheme, we find that the 
lowest order new operator is $\sim a \epsilon^2 \bar\psi\psi$.
This is, however, of higher order than we consider here.\footnote{%
Furthermore, when mapped to the chiral Lagrangian, it leads to contributions
which can be absorbed by making the untwisted mass $m$ have a weak dependence
on $\epsilon$. Thus it does not lead to new phases, but only to a small
distortion of the phase diagram.}
All other operators are of yet higher order.
Thus, at the order we work,
non-degeneracy only enters our calculation through the continuum $\ell_7$ term.
The LO potential thus becomes
\begin{align}
{\mathcal{V}_{a^2,\ell_7}}=&
-\frac{f^2}{4} \tr(\chi^\dagger \Sigma + \Sigma^\dagger \chi)
- W' [\tr(A^\dagger \Sigma + \Sigma^\dagger   A)]^2
+\frac{\ell_7}{16}[\tr(\chi^\dagger \Sigma - \Sigma^\dagger \chi)]^2
\,.
\label{eq:VA2ell7}
\end{align}
We stress that it is self-consistent to determine
the vacuum structure and pion masses from a tree-level analysis of
${\cal V}_{a^2,\ell_7}$ since loop effects only come in at
${\cal O}(m^2,ma^2,a^4)$.

\begin{figure}[tb!]
\centering
\begin{subfigure}{0.49\textwidth}
\includegraphics[scale=.29]{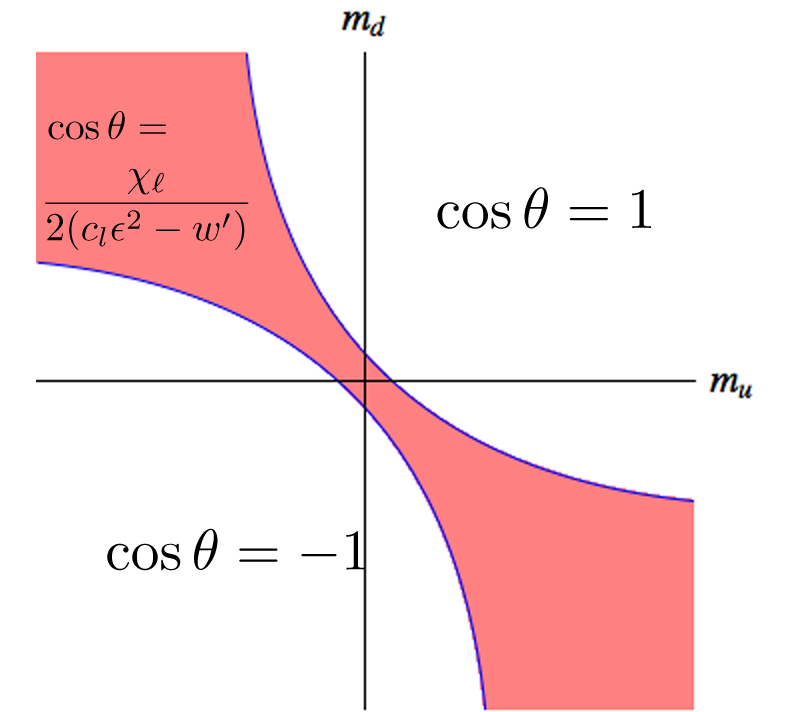}
\caption{\label{fig:NLOAoki} Aoki scenario ($w'<0$).}
\end{subfigure}
\begin{subfigure}{0.49\textwidth}
\includegraphics[scale=.29]{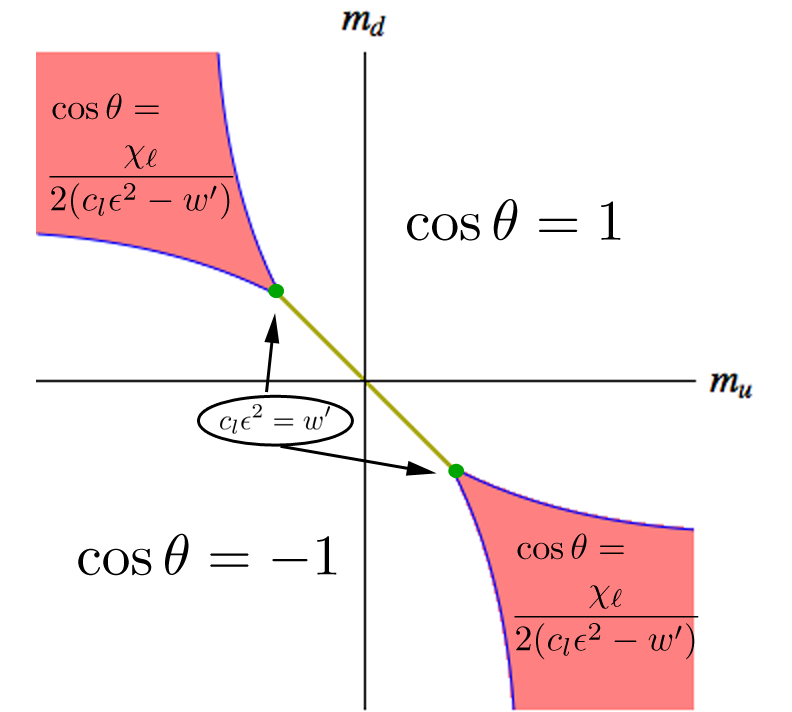}
\caption{\label{fig:NLOFirst} First-order scenario ($w'>0$).}
\end{subfigure}
\caption{\centering Phase diagrams including effects of both
discretization and non-degeneracy.
Blue (yellow) lines indicate second (first) order transitions.
Equations for the positions of phase transition lines are given in the text.}
\label{fig:untwistPhases}
\end{figure}

In terms of the parameters of $\langle\Sigma\rangle$, the potential is now
given by
\begin{equation}
- \frac{\mathcal{V}_{a^2,\ell_7}}{f^2} 
= \chi_\ell\cos{\theta} +c_\ell \epsilon^2 n_3^2
\sin^2{\theta} +w' \cos^2{\theta}\,.
\end{equation}
As before, we can set $n_3=1$ without loss of generality.
The stationary points are at $\cos{\theta}=\pm1$ and
\begin{equation}
\cos{\theta} = \frac{\chi_\ell}{2(c_\ell \epsilon^2-w')}\,.
\label{eq:costhetares}
\end{equation}
The latter minimizes the potential if $c_\ell\epsilon^2-w'>0$ 
and is valid for $|\cos\theta|\le 1$.
This results in the phase diagrams of
Figs.~\ref{fig:NLOAoki} and \ref{fig:NLOFirst} for
$w'<0$ and $w'>0$, respectively.
In the former case, corresponding to the Aoki phase for degenerate
quarks, the second-order transition lines lie at
\begin{equation}
\chi_\ell= \pm 2 (c_\ell \epsilon^2-w')
\label{eq:secondboundary}
\,.
\end{equation}
Thus the width of the phase grows as $|\epsilon|$ increases.
Furthermore, comparing to Fig.~\ref{fig:AokiLO}, we see that the
continuum CP-violating phase and the Aoki phase are continuously 
connected.\footnote{%
This result is in agreement with Creutz' conjecture~\cite{Creutz:2014em}.}
The only subtlety in this connection is that the condensate
definitely points in the $n_3$ direction for $\epsilon\ne 0$
(i.e. the direction picked out by the non-degenerate part of the mass term),
whereas for $\epsilon=0$ the direction is arbitrary.

In the first-order scenario, Fig.~\ref{fig:NLOFirst},
the first-order transition along the $m_u=-m_d$ line
weakens as $|\epsilon|$ increases,
until, at $c_\ell \epsilon^2=w'$, the CP-violating phase appears.
The second-order transition lines are then given by
$|\chi_\ell|=2 (c_\ell \epsilon^2-w')$, i.e. by the same
equation as in the Aoki scenario.

\begin{figure}[tb!]
\centering
 \begin{subfigure}{0.49\textwidth}
\includegraphics[scale=.3]{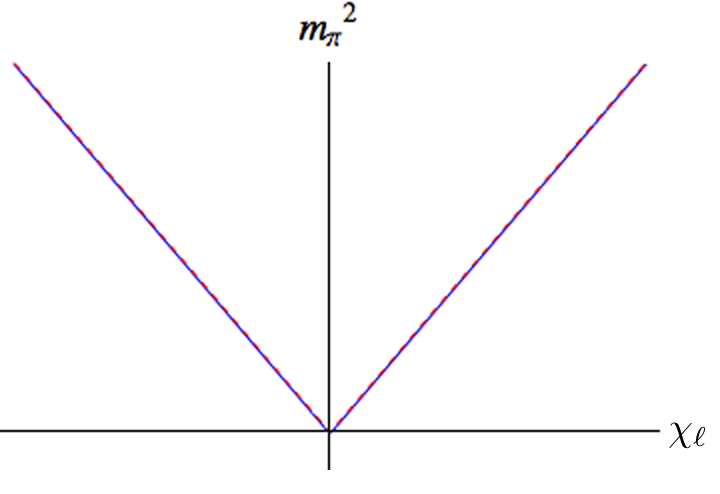}
\caption{$w'=c_\ell\epsilon^2=0$}
\end{subfigure}
 \begin{subfigure}{0.49\textwidth}
\includegraphics[scale=.3]{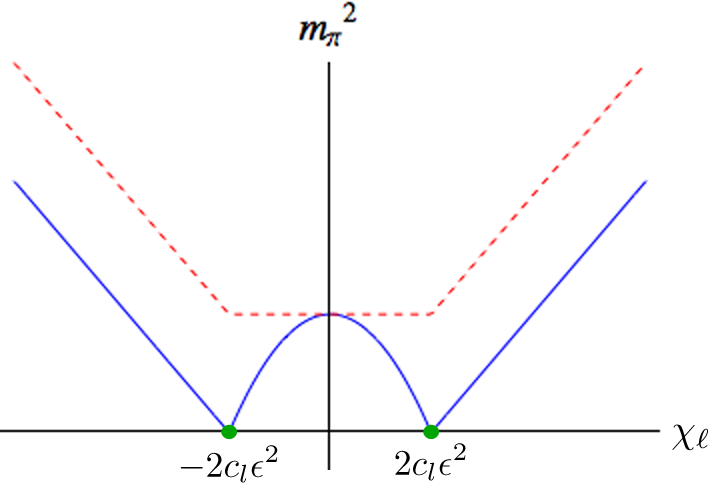}
\caption{$w'=0$, $c_\ell\epsilon^2>0$}
\end{subfigure}

 \begin{subfigure}{0.49\textwidth}
\includegraphics[scale=.3]{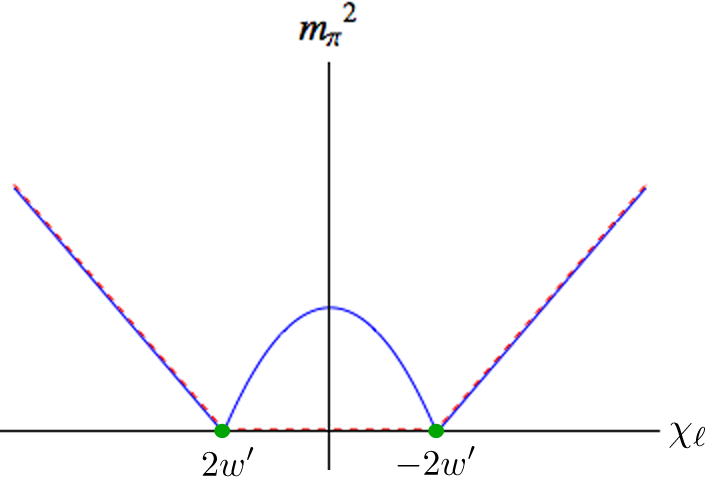}
\caption{$w'<0$, $c_\ell\epsilon^2= 0$}
\end{subfigure}
 \begin{subfigure}{0.49\textwidth}
\includegraphics[scale=.3]{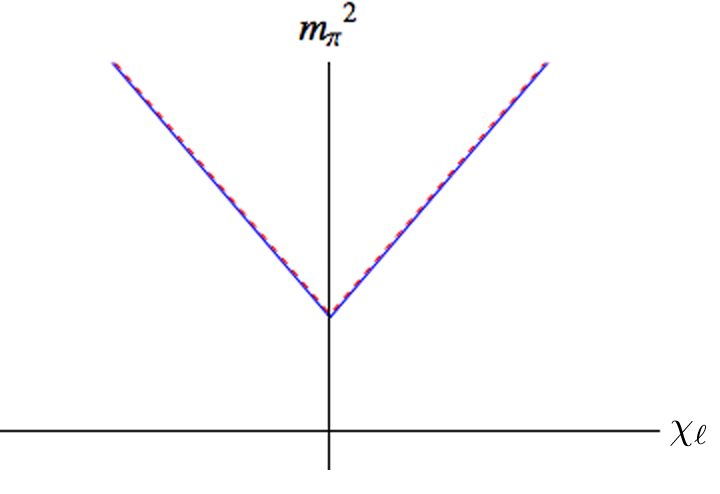}
\caption{$w'>0$, $c_\ell\epsilon^2=0$}
\end{subfigure}

 \begin{subfigure}{0.49\textwidth}
\includegraphics[scale=.3]{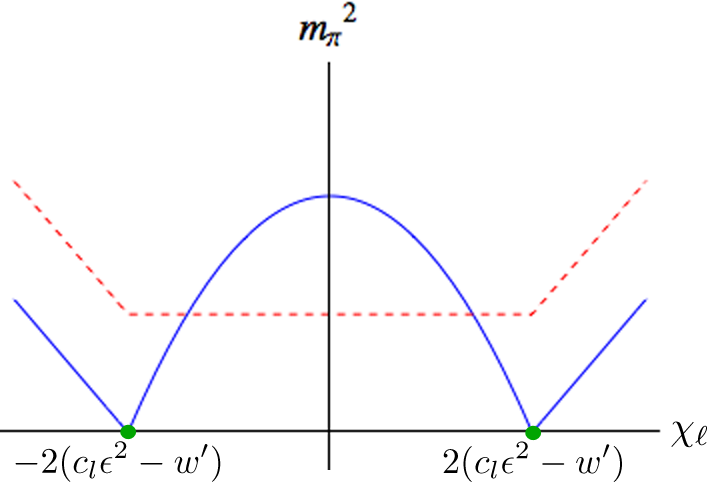}
\caption{$w'<0$, $c_\ell\epsilon^2>0$}
\end{subfigure}
 \begin{subfigure}{0.49\textwidth}
\includegraphics[scale=.3]{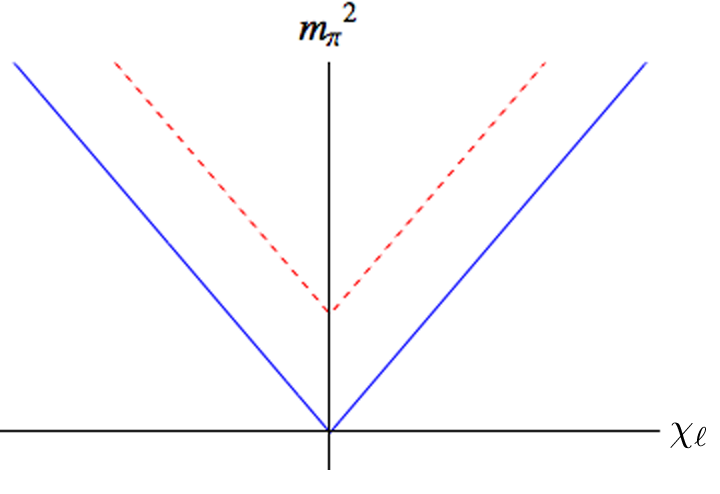}
\caption{ $c_\ell\epsilon^2=w'>0$}
\end{subfigure}
 \begin{subfigure}{0.49\textwidth}
\includegraphics[scale=.3]{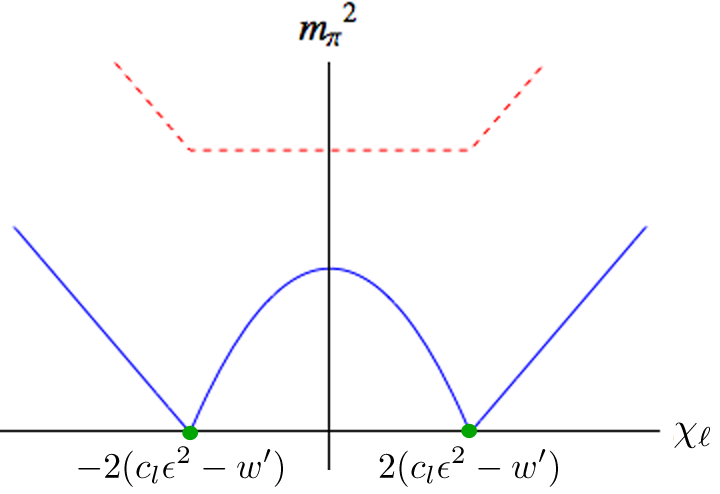}
\caption{$c_\ell\epsilon^2>w'>0$}
\end{subfigure}
\caption{\label{fig:UnTwistPiMasses} 
Pion masses for untwisted Wilson
fermions including the effects of both discretization ($w'\ne 0$)
and non-degeneracy ($\epsilon\ne 0$).
$m_{\pi^0}^2$ is shown by solid (blue) lines,
$m_{\pi^\pm}^2$ by dashed (red) lines.
Explicit expressions for the masses 
are given in the text.
Vertical scales differ between the figures.}
\end{figure}

We next calculate the pion masses throughout the phase plane,
expanding about the vacuum as
\begin{equation}
\Sigma= \exp(i\theta \tau_3) \exp(i\vec\pi\cdot\vec\tau/f)\,.
\end{equation}
Outside the CP-violating phase, we find
\begin{align}
m_{\pi^0}^2&= |\chi_\ell| - 2 (c_\ell \epsilon^2-w')\,,
\label{eq:mpi0untwistLO}
\\
m_{\pi^\pm}^2&= m_{\pi^0}^2 + 2 c_\ell \epsilon^2\,.
\label{eq:mpipuntwistLO}
\end{align}
while within the CP-violating phase we have
\begin{align}
m_{\pi^0}^2&=
2( c_\ell \epsilon ^2  -w') \sin^2\theta
\label{eq:mpi0CPuntwistLO}
\\
m_{\pi^\pm}^2&= 2c_\ell \epsilon^2\,,
\label{eq:mpipCPuntwistLO}
\end{align}
where $\theta$ is given in Eq.~(\ref{eq:costhetares}).
These results are plotted versus $\chi_\ell$ for various characteristic
choices of $\epsilon$ and $w'$ in Fig.~\ref{fig:UnTwistPiMasses}.

Figures~\ref{fig:UnTwistPiMasses}a and b show the continuum results
for degenerate and non-degenerate masses, respectively.
The neutral pion mass vanishes along the second-order transition line, 
as expected. 
The full degeneracy at $\chi_\ell=0$ is due to the fact that the
theory regains flavor symmetry (with $\theta_{\rm QCD}=\pi$)
at this point.
A characteristic feature of the spectrum at this order
is that the charged pion mass
is independent of $\chi_\ell$ within the CP-violating phase.
This holds also when discretization errors are included.

Figures~\ref{fig:UnTwistPiMasses}c and d show the spectrum
for degenerate quarks with discretization errors included,
respectively for the Aoki and first-order scenarios,
reproducing the results of Ref.~\cite{Sharpe:1998xm}.

Our new results are those of Figs.~\ref{fig:UnTwistPiMasses}e-g,
which include the effects of both discretization errors and
non-degeneracy. In this case the charged and neutral pion masses
differ in general.
Figure~\ref{fig:UnTwistPiMasses}e
shows the behavior in the Aoki scenario,
where $m_{\pi^0}$ vanishes on the phase transition lines, and 
rises above $m_{\pi^\pm}$ in the central region of the
CP-violating phase. There are thus two values of
$\chi_\ell$ where all pions are degenerate, but these are
accidental degeneracies and not indicative of any symmetry.
For the first-order scenario
Fig.~\ref{fig:UnTwistPiMasses}f shows the spectrum when $\epsilon$
is chosen so that the plot passes through the end-point of the
second-order transition line, while
Fig.~\ref{fig:UnTwistPiMasses}g shows what happens as one moves
through the CP-violating phase. In this case, there are no degenerate
points.

Simulations using Wilson-like fermions at physical masses,
including isospin breaking, have recently 
begun~\cite{Borsanyi:2014jba}.
What is the significance of our results for such simulations?
The main issue is whether discretization effects can move the
CP-violating phase such that it lies closer to,
or even includes, the physical point.
Clearly one wants to avoid simulating in this phase, since it
has a different vacuum structure from the continuum theory.
But even lying close to a second-order transition could lead to
algorithmic issues due to critical slowing down.
What we have found is that the phase does move closer
to the physical point in the Aoki scenario, Fig.~\ref{fig:NLOAoki}.
In this scenario, the CP-violating 
phase now includes a region of positive quark masses.
On the other hand, for the first-order scenario, 
discretization effects move the CP-violating phase 
away from the physical point.
A positive aspect of our
results is that discretization errors lead only to a overall shift
in pion masses (outside of the CP-violating phase), so that
the difference $m_{\pi^\pm}^2-m_{\pi^0}^2$ takes its
continuum value $2c_\ell \epsilon^2$ in both scenarios.

\clearpage

\section{Twisted-mass fermions at maximal twist}
\label{sec:twist}

In this section we extend the previous analysis to
twisted-mass fermions~\cite{Frezzotti:2000nk} at maximal twist.
Such fermions have the important practical property of automatic
${\cal O}(a)$ improvement~\cite{Frezzotti:2003ni}. 
They are being used to simulate QCD with quarks at or near their physical 
masses~\cite{Abdel-Rehim:2013yaa,Carrasco:2014cwa}, 
and isospin breaking is now being included~\cite{deDivitiis:2013xla}.
The main question we address here is the same as for untwisted fermions:
How do discretization effects change the continuum phase structure
and pion masses?

In the continuum, twisted mass fermions are obtained by a non-anomalous
axial rotation,
\begin{equation}
\mathcal{L}_{\rm QCD} = 
\overline{\psi} (\slashed{D}+m_\ell +\epsilon_\ell \tau_3) \psi 
\rightarrow 
\overline{\psi} (\slashed{D}+ m_\ell e^{i\gamma_5 \tau_1 \omega} 
+\epsilon_\ell \tau_3)\psi
=
\overline{\psi} (\slashed{D}+ m + i \gamma_5 \tau_1 \mu
+\epsilon_\ell \tau_3)\psi
\,,
\end{equation}
with $m_\ell=(m_u\!+\!m_d)/2$, $\epsilon_\ell=(m_u\!-\!m_d)/2$,
$m=m_\ell\cos\omega$, $\mu=m_\ell\sin\omega$, and $\omega$ the twist angle.
Conventionally, $m$ is called the untwisted (average) mass
and $\mu$ the twisted (average) mass.
Choosing the twist in a direction orthogonal to $\tau_3$
leaves the $\epsilon_\ell$ term unchanged.
In the continuum this is a convenience, but not a necessity.
Once one discretizes $\slashed{D}$ with a Wilson term, however,
it is mandatory to twist in a direction orthogonal to $\tau_3$ if
one wants to keep the fermion determinant 
real~\cite{Frezzotti:2003xj}.\footnote{%
In Ref.~\cite{deDivitiis:2013xla}, which studies twisted-mass
non-degenerate fermions, the twist is chosen in the $\tau_3$
direction. This leads to a complex fermion determinant,
which is avoided in practice by perturbing at linear order
around the isospin-symmetric theory.
Because the twist is in the $\tau_3$ direction, our 
present results do not apply to these simulations.
We will discuss the generalization to $\tau_3$ twist
(along with the inclusion of electromagnetism) in
an upcoming work~\cite{inprep}.}
By convention, this direction is chosen to be $\tau_1$.
The rescaled mass matrix that enters \chpt\ is now
\begin{equation}
\chi=\chi_\ell e^{i \tau_1 \omega} +\epsilon \tau_3 = 
\chi_\ell \cos{\omega} \mathbb{1} \
+i \chi_\ell \sin{\omega} \tau_1  +\epsilon \tau_3 = 
\mhat \mathbb{1} +i \muhat\tau_1 +\epsilon \tau_3\,,
\label{eq:twistedchi}
\end{equation}
and is no longer hermitian.
Here we have defined  
\begin{equation}
\mhat\equiv 2B_0 m=\chi_\ell \cos\omega \ \ {\rm and}\ \
\muhat\equiv 2B_0 \mu=\chi_\ell\sin\omega
\end{equation}
following Ref.~\cite{Sharpe:2004ps}.

To determine the effective chiral theory for
twisted-mass lattice QCD the first step is to determine 
the additional operators in the Symanzik Lagrangian that are
induced by twisting.
As in the untwisted case, the form of the allowed operators can be
obtained from the analysis of Ref.~\cite{WalkerLoud:2005bt}, 
which includes both twist and non-degeneracy.
In fact, since $\muhat^2$ is smaller than $\epsilon^2$ in our power-counting,
the inclusion of twist does not change the result for the untwisted case,
namely that the lowest order new operator is $\sim a\epsilon^2$ and of higher
order than we are working.
Thus at LO the extension \chpt\ to include twist 
and discretization errors is accomplished
by simply using the twisted $\chi$ of Eq.~(\ref{eq:twistedchi})
in the potential ${\cal V}_{a^2,\ell_7}$ of Eq.~(\ref{eq:VA2ell7}).

Using our standard parametrization of $\langle\Sigma\rangle$ this gives
\begin{equation}
- \frac{\mathcal{V}_{a^2,\ell_7}}{f^2} = 
\mhat\cos{\theta} +\muhat n_1 \sin{\theta}
+c_\ell \epsilon^2 n_3^2 \sin^2{\theta} +w'\cos^2{\theta}\,.
\label{eq:Vtwist}
\end{equation}
We focus in this section on the case of maximal twist, $\mhat=0$,
where simple analytic results can be obtained. 
Even with this simplification, we note that there is competition
between terms in three directions in $\Sigma$:
the twist direction $n_1$, the non-degeneracy direction $n_3$, 
and the identity direction ($w'$ term).
Thus we can expect a more complicated phase structure than
for untwisted Wilson fermions.
Furthermore, since non-degenerate twisted-mass quarks
completely break the continuous SU(2) flavor symmetry,
we expect, in general, that all three pion masses will differ.

\begin{figure}[tb!]
\centering
 \begin{subfigure}{0.99\textwidth}
 \centering
 \includegraphics[scale=.4]{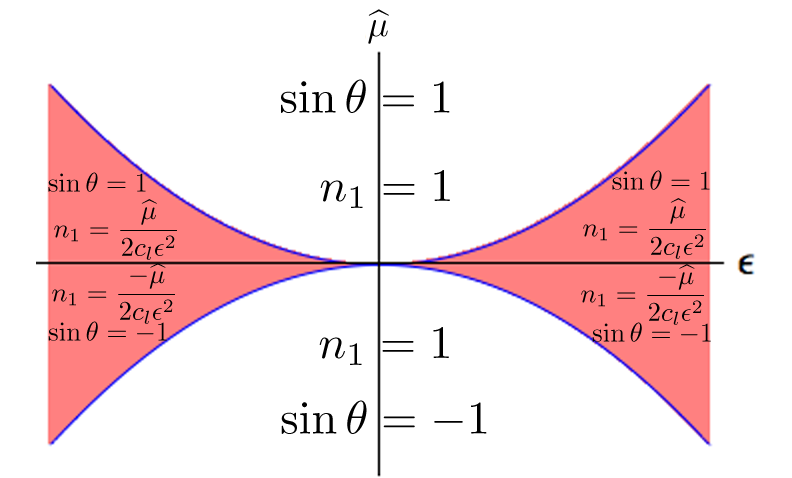}
\caption{Aoki scenario or continuum ($w'\leq0$)}
\end{subfigure}

 \begin{subfigure}{0.99\textwidth}
 \centering
 \includegraphics[scale=.4]{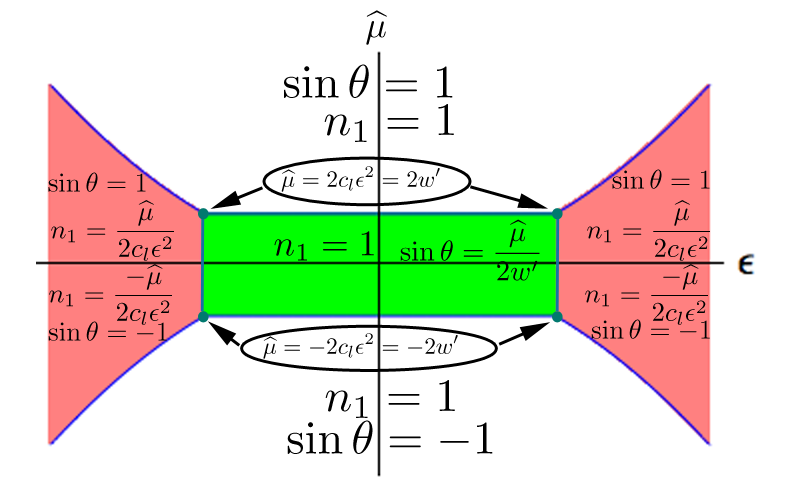}
\caption{First-order scenario ($w'>0$)}
\end{subfigure}
\caption{\label{fig:maxTwistPhases} 
Phase diagrams at maximum twist ($\mhat=0$).}
\end{figure}

We find the phase diagrams shown in Fig.~\ref{fig:maxTwistPhases}.
Note that we are now plotting the average mass along the vertical
axis and the difference horizontally. 
We do this because $\muhat$ and $\epsilon$
are proportional to parameters that enter the twisted-mass lattice action.
To compare to the earlier plots, one should rotate those of
Fig.~\ref{fig:maxTwistPhases} by $45^\circ$ in a clockwise direction.
We see that, at maximal twist, it is the Aoki scenario
which is preferred, in the sense that the CP-violating phase does
not move closer to the physical point. Indeed, the phase diagram
in this scenario is identical to that in the continuum, Fig.~\ref{fig:NLOell7},
with the replacement $\chi_\ell\to\muhat$.
In the first-order scenario, by contrast, there is an additional
phase (colored green in Fig.~\ref{fig:maxTwistPhases}b)
which brings lattice artifacts closer to the physical point.
Thus the relative merits of the two scenarios are interchanged
compared to the untwisted case.

To understand the phase diagrams we first recall the
result for the degenerate case, $\epsilon=0$, which has been studied in 
Refs.~\cite{Munster:2004am,Scorzato:2004da,Sharpe:2004ps}. 
These works find, for large $|\muhat|$, 
that the condensate is aligned with the twist, i.e. $n_1=1$
and $\sin\theta={\rm sign}(\muhat)$. This is as in the continuum.
In the Aoki scenario ($w'<0$), this alignment holds for all $\muhat$,
and there is a first-order transition at $\muhat=0$ where $\sin\theta$ 
changes sign.
In the first-order scenario ($w'>0$), there are second-order transitions at
the two points $\muhat=\pm 2w'$, at which one of the pion masses vanishes.
For $|\muhat|< 2w'$ the condensate smoothly rotates within the group manifold
with $\sin\theta=\muhat/(2w')$. 
These features are reproduced by our results along the
vertical axes in Fig.~\ref{fig:maxTwistPhases}.

We now explain how these results are generalized to $\epsilon\ne 0$.
We first observe that we can set $n_2=0$. This is because, for
any choice of $n_1$, the $c_\ell$ term in
Eq.~(\ref{eq:Vtwist}) (with $c_\ell>0$) will be minimized when $n_3^2$ is maximized,
i.e. with $n_3^2=1-n_1^2$.
Thus there are only two independent variables, $\theta$ and $n_1$.
Since $n_1$ satisfies $|n_1|\le 1$, we parametrize it as $n_1=\cos\varphi_1$.
Since $\langle\Sigma\rangle$ is invariant when $\theta$ and $\vec n$
change sign, we need only consider $n_1\ge 0$, i.e. $0\le\varphi_1\le\pi/2$.
The stationary points are obtained from simultaneously solving
\begin{align}
\frac{\partial \mathcal{V}_{a^2,\ell_7}}{\partial \theta} &\propto 
\cos\theta\left[\muhat \cos\varphi_1
+  2\sin{\theta}(\sin^2\varphi_1c_\ell\epsilon^2-w')\right]
=0\,,
\label{eq:partialVtheta}
\\
\frac{\partial \mathcal{V}_{a^2,\ell_7}}{\partial \varphi_1} &\propto
\sin\theta\sin\varphi_1\left[\muhat  -2\sin{\theta}\cos\varphi_1c_\ell\epsilon^2 \right]
=0\,.
\label{eq:partialVtheta1}
\end{align}
The solutions are 
\begin{enumerate}
\item
$\cos\theta=0$ (so that $\sin\theta=\pm1$) together with
$\sin\varphi_1=0$ (so that $n_1=1$).
In these cases
${\mathcal{V}_{a^2,\ell_7}}/{f^2} = \mp \muhat$, so that the solution
with the lowest energy is that with $\sin\theta={\rm sign}(\muhat)$,
giving ${\mathcal{V}_{a^2,\ell_7}}/{f^2} = - |\muhat|$.
\item
$\sin\theta={\rm sign}(\muhat)$
and $n_1=\cos\varphi_1= {|\muhat|}/({2c_\ell \epsilon^2})$
so that
${\mathcal{V}_{a^2,\ell_7}}/{f^2} = - {\muhat^2}/({4 c_\ell \epsilon^2})
- c_\ell \epsilon^2$.
This is only valid when $n_1\le 1$, i.e. $|\muhat|\le 2 c_\ell \epsilon^2$.
There are two degenerate solutions, with $n_3=\pm \sin\varphi_1$.
\item
$\sin\theta=\muhat/(2w')$ and $\varphi_1=0$ (implying $n_1=1$) so
that 
${\mathcal{V}_{a^2,\ell_7}}/{f^2} =- \muhat^2/(4 w') - w'$.
This is only valid when $|\muhat|\le 2 w'$.
There are two degenerate solutions, with opposite signs of $\cos\theta$.
\item
$\cos\theta=\pm1$ and $\muhat n_1=0$, 
so that $\mathcal{V}_{a^2,\ell_7}/f^2=- w'$.
This never has lower energy than the third solution and can be ignored.
\end{enumerate}
The first solution is the continuum one discussed above.
The second has lower energy than the first where it is valid, and 
goes over to the CP-violating phase when $w'=0$.
The third solution is relevant only for $w'>0$, in which case it has the
lowest energy when $c_\ell \epsilon^2 < w'$. The condensate in this
phase is independent of $\epsilon$.
These considerations lead to the phase diagrams shown in 
Fig.~\ref{fig:maxTwistPhases}.
The potential is continuous throughout the phase planes,
as is the condensate except at the junction between the
central (green colored) phase and the CP-violating phase in
Fig.~\ref{fig:maxTwistPhases}b.
Thus we expect the transitions to be of second order.

We calculate pion masses using the parametrization
\begin{equation}
\Sigma= \exp(i\theta\hat{n} \cdot \vec{\tau}/2) 
\exp(i\vec \pi\cdot \vec\tau/f)
\exp(i\theta\hat{n} \cdot \vec{\tau}/2)\,, \qquad
\left[\langle\Sigma\rangle= \exp(i\theta\hat{n} \cdot \vec{\tau}) \right]\,.
\label{eq:axialparam}
\end{equation}
Here we are using an axial transformation to rotate from the
twisted basis to the physical basis, which ensures, in the continuum,
that the pion fields have physical flavors~\cite{Sharpe:2004ny}.
In the continuum-like phase (uncolored in the figures), which lies in the
regions $|\muhat|\ge {\rm max}(2c_\ell\epsilon^2,2w')$, we find
\begin{align}
&m_{{\pi}_1}^2= |\muhat|-2 w'\,, 
& m_{{\pi}_2}^2= |\muhat|\,,
&& m_{{\pi}_3}^2= |\muhat|-2 c_\ell\epsilon^2\,.
\label{eq:mpiwhite}
\end{align}
These results are consistent with those of Ref.~\cite{Munster:2006yr},
where a \chpt\ calculation in this phase
is carried out using the different power-counting $m\gtrsim a$.
Various aspects of these results are noteworthy.
First, all three masses differ. This is expected since flavor symmetry
is completely broken. 
Second, the charged pions are not mass eigenstates; instead, 
the eigenstates are $\pi_{1,2}$ and the neutral pion.
These two points were also noted in Ref.~\cite{Munster:2006yr}.
Third, one of the pion masses vanishes at each of the phase boundaries:
$m_{\pi_3}$ at the boundary with the CP-violating (pink colored) phase,
and $m_{\pi_1}$ at the boundary with the central (green colored) phase
in the first-order scenario.\footnote{%
In the degenerate case ($\epsilon=0$) 
Refs.~\cite{Munster:2004am,Scorzato:2004da,Sharpe:2004ps} find
that it is $m_{\pi_3}$ which vanishes at $|\muhat|=2w'$, rather than
$m_{\pi_1}$. This difference arises because we twist in the $\tau_1$
direction rather than the $\tau_3$ direction used in
Refs.~\cite{Munster:2004am,Scorzato:2004da,Sharpe:2004ps}.}
This is expected since these are continuous transitions at which
a $Z_2$ symmetry is broken ($\theta \to -\theta$ for the ``green phase''
and $n_3\to -n_3$ for the CP-violating phase).
Finally, in the first-order scenario, there are four tricritical points
at which {\em both} $m_{\pi_3}$ and $m_{\pi_1}$ vanish.
These occur where all three phases meet, i.e. at
$|\muhat|=2c_\ell\epsilon^2=2w'$.

In the central (green) phase we find
\begin{align}
&m_{{\pi}_1}^2= 2w'-\frac{\muhat^2}{2w'}\,, 
& m_{{\pi}_2}^2= 2w'\,,
&& m_{{\pi}_3}^2= 2w'-2 c_\ell\epsilon^2\,.
\label{eq:mpigreen}
\end{align}
Thus $m_{\pi_2}$ and $m_{\pi_3}$ are independent of $\muhat$ within
this phase. These results agree with those in the normal
phase, Eq.~(\ref{eq:mpiwhite}), at the boundaries.
They also show that $m_{\pi_3}$ vanishes at the borders with
the CP-violating (pink) phases ($c_\ell\epsilon^2=w'$).

In the CP-violating phase there is mixing between $\pi_1$
and $\pi_3$, with the mass eigenvectors being
\begin{equation}
\tilde\pi_1= n_1 \pi_1 + n_3 \pi_3\ \ {\rm and}\ \ 
\tilde\pi_3= -n_3 \pi_1 + n_1 \pi_3\,,
\end{equation}
where we recall that $n_1=\muhat/(2c_\ell \epsilon^2)$
and $n_3=\sqrt{1-n_1^2}$.
The masses are
\begin{align}
&m_{{\tilde\pi}_1}^2= 2 c_\ell\epsilon^2- 2w'\,,
& m_{{\pi}_2}^2= 2 c_\ell\epsilon^2\,,
&& m_{{\tilde\pi}_3}^2= 
2 c_\ell\epsilon^2-\frac{\muhat^2}{2 c_\ell\epsilon^2}\,.
\label{eq:mpipink}
\end{align}
Note that $m_{{\tilde\pi}_1}$ and $m_{\pi_2}$ are independent of $\muhat$,
while the $\tilde\pi_3$ mass vanishes along the boundaries 
with the standard phases. The latter result is consistent with
the results above because, on these boundaries $|n_1|=1$
and so $\tilde\pi_3 =\pm \pi_3$.

A puzzling feature of these results is what happens at the
boundaries between the central (green) and CP-violating (pink) phases.
According to Eq.~(\ref{eq:mpigreen}) it is the mass of $\pi_3$
which vanishes there, while Eq.~(\ref{eq:mpipink}) has the mass
of $\tilde\pi_1$ vanishing. These appear to be different particles.
This is related to a second puzzle, namely 
that the condensate is discontinuous across the boundary
(which lies at $w'=c_\ell \epsilon^2$):
\begin{equation}
\langle\Sigma\rangle_{\rm Green}^{\rm Boundary}
= i \frac{\muhat}{2w'} \tau_1
\pm \sqrt{1-\frac{\muhat^2}{4w'^2}}\mathbb 1
\ \ {\rm vs.}\ \
\langle\Sigma\rangle_{\rm Pink}^{\rm Boundary}
= i \frac{\muhat}{2w'} \tau_1
\pm i \sqrt{1-\frac{\muhat^2}{4w'^2}} \tau_3
\,.
\label{eq:condjump}
\end{equation}
Here the $\pm$ signs correspond to the two choices
of vacuum state on each side.
This situation can be understood by noting that, at the boundary,
the vacuum manifold expands to a line which
includes all four values of the condensate given in Eq.~(\ref{eq:condjump}):
\begin{equation}
\langle\Sigma\rangle = i \frac{\muhat}{2w'} \tau_1
+ \sqrt{1-\frac{\muhat^2}{4w'^2}}
(\cos\phi+i \tau_3 \sin\phi)\,,
\end{equation}
where $\phi$ is arbitrary.
The presence of this flat direction is the reason that one pion is massless, 
since there is no breaking of a $Z_2$ symmetry to explain the masslessness.
The orientation of the flat direction,
which is the direction of the massless pion, depends on the
position along this vacuum manifold, and thus is different
on the two sides of the transition.
In this way to two puzzles above are simultaneously explained.

\begin{figure}[tb!]
\centering
 \begin{subfigure}{0.49\textwidth}
\includegraphics[scale=.28]{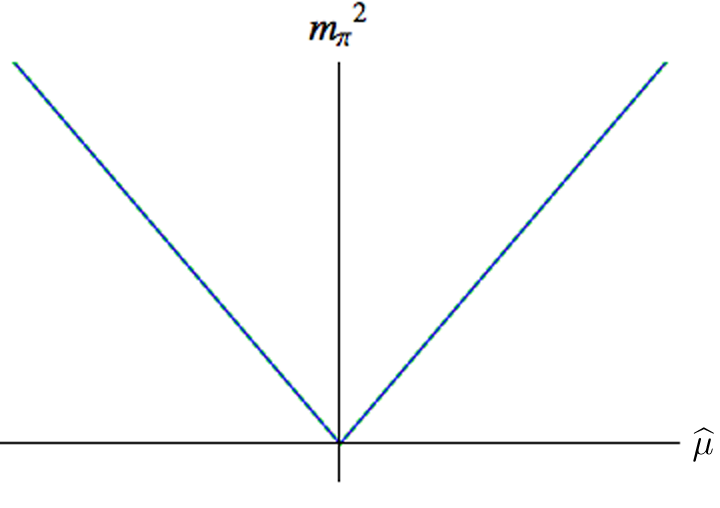}
\caption{ $w'=c_\ell\epsilon^2=0$}
\end{subfigure}
 \begin{subfigure}{0.49\textwidth}
\includegraphics[scale=.28]{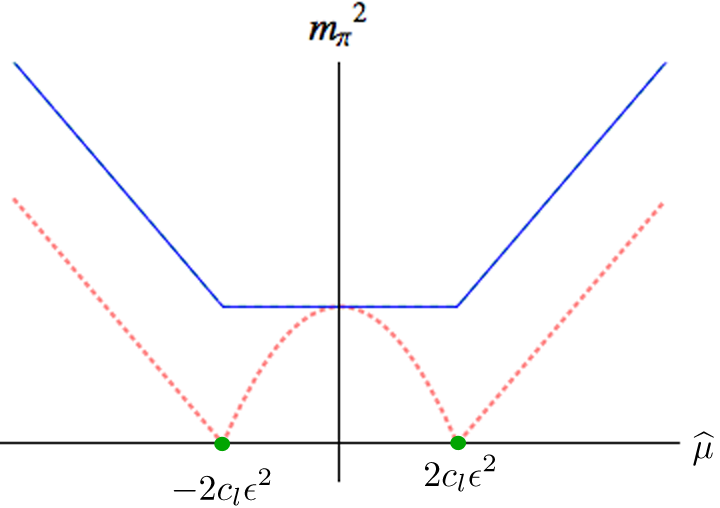}
\caption{$w'=0$, $c_\ell\epsilon^2>0$} 
\end{subfigure}

 \begin{subfigure}{0.49\textwidth}
\includegraphics[scale=.28]{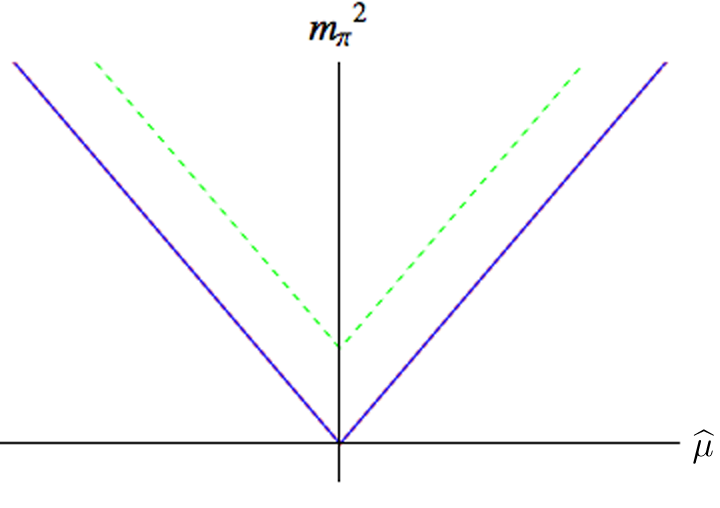}
\caption{$w'<0$, $c_\ell\epsilon^2= 0$} 
\end{subfigure}
 \begin{subfigure}{0.49\textwidth}
\includegraphics[scale=.28]{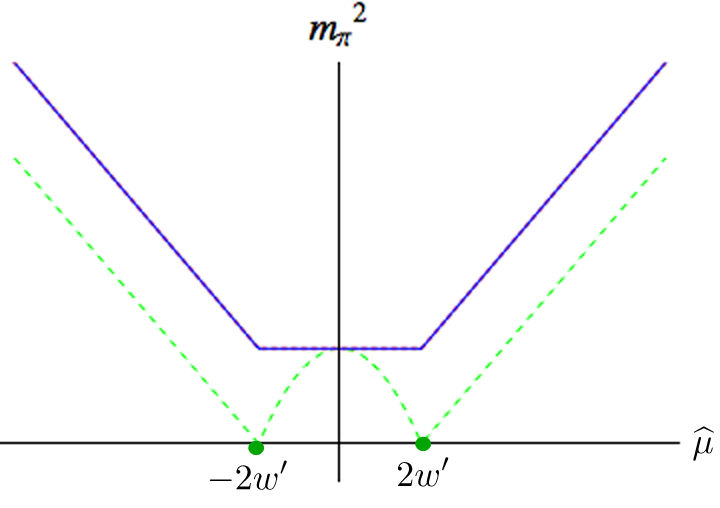}
\caption{$w'>0$, $c_\ell\epsilon^2= 0$} 
\end{subfigure}

 \begin{subfigure}{0.49\textwidth}
\includegraphics[scale=.28]{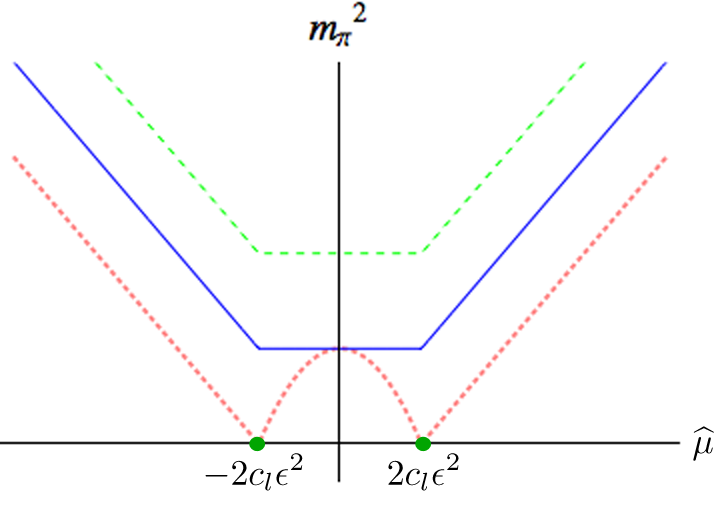}
\caption{$w'<0$, $c_\ell\epsilon^2>0$} 
\end{subfigure}

 \begin{subfigure}{0.49\textwidth}
\includegraphics[scale=.28]{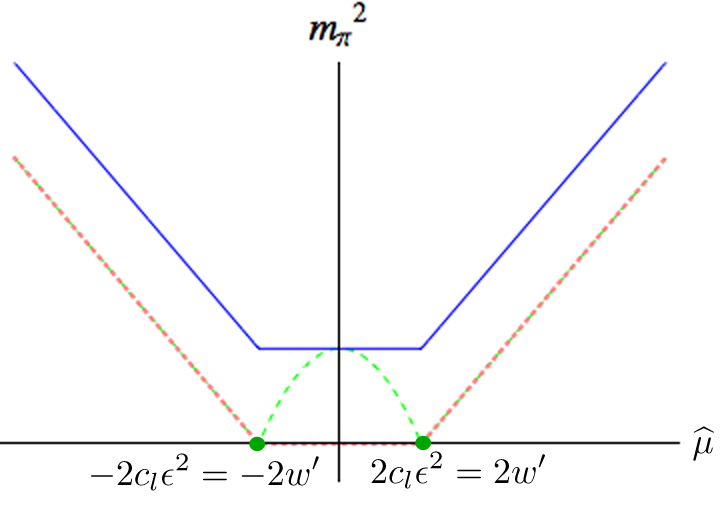}
\caption{$w'>0$, $c_\ell\epsilon^2=w'$} 
\end{subfigure}
 \begin{subfigure}{0.49\textwidth}
\includegraphics[scale=.28]{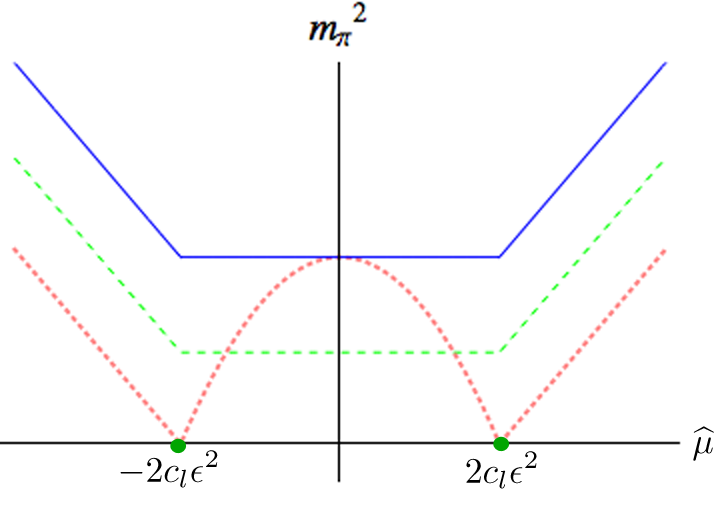}
\caption{$w'>0$, $c_\ell\epsilon^2>w'$} 
\end{subfigure}
\caption{\label{fig:MaxTwistPiMasses} 
Pion masses for maximally twisted
fermions including the effects of both discretization ($w'\ne 0$)
and non-degeneracy ($\epsilon\ne 0$).
$m_{\pi_2}^2$ is shown by solid (blue) lines,
$m_{\pi_3}^2$ (and $m_{\tilde\pi_3}^2$) by dotted (red) lines and
$m_{\pi_1}^2$ (and $m_{\tilde\pi_1}^2$) by dashed (green) lines.
Not all lines are visible in some figures due to degeneracies.
Mixing of pions occurs only within the CP-violating phase in
Figs. (e) and (g).
Explicit expressions for masses and mixing
are given in the text.
Vertical scales differ between the figures.}
\end{figure}

Results for pion masses are plotted in Fig.~\ref{fig:MaxTwistPiMasses}.
We choose the same parameters for the plots as for the untwisted
case, Fig.~\ref{fig:UnTwistPiMasses}, so as to allow a clear comparison.
The figures illustrate the discussion given above.

\section{Arbitrary Twist}
\label{sec:arbtwist}

In this section we give a brief discussion of the phase diagram
at arbitrary twist. 
This allows us to understand how the phase diagrams presented
above for untwisted and maximally-twisted quarks are related to
one another. We focus on the phase diagram, and in particular, the
position of the critical manifold where one or more pions are massless.

For arbitrary twist, the potential is given in Eq.~(\ref{eq:Vtwist}).
As before, minimization leads to $n_2=0$, so the potential depends only on
$\theta$ and $\varphi_1$ (defined by $\cos\varphi_1=n_1$).
The equations for stationary points are
\begin{equation}
- \mhat \sin{\theta} 
+ \cos\theta\left[\muhat \cos\varphi_1
+  2\sin{\theta}(c_\ell\epsilon^2\sin^2\!\varphi_1-w')\right] = 0\,,
\label{eq:partialVthetab}
\end{equation}
and Eq.~(\ref{eq:partialVtheta1}).
We focus on the case when both $\mhat$ and $\muhat$ are non-zero,
since the special cases when one of these vanish have been discussed above.

When $|\muhat|,|\mhat|\gg c_\ell\epsilon^2,|w'|$ the solution
which minimizes the potential has
\begin{equation}
n_1=\cos\varphi_1=1,\ \  n_3=\sin\varphi_1,\ \  
\tan{\theta}\approx\frac{\muhat}{\mhat}\,.
\end{equation}
The last equation becomes an equality in the continuum limit,
and simply describes how the condensate twists to compensate the
twist in the mass. Discretization errors (here proportional to $w'$) 
lead to a small deviation in $\theta$ from this continuum result.
We do not quote the analytic form as it is not illuminating.
In fact, the result for $\theta$ turns out to be independent 
of the non-degeneracy $\epsilon$, so the results for the condensate given 
for the degenerate theory in 
Refs.~\cite{Munster:2004am,Scorzato:2004da,Sharpe:2004ps} remain valid
in this phase. This phase is the extension of the ``uncolored'' phases 
in Figs.~\ref{fig:untwistPhases} and \ref{fig:maxTwistPhases}
to arbitrary twist.
At a general position in this phase, the mass eigenstates are
$\pi_1$, $\pi_2$ and $\pi_3$ 
[using the parametrization of Eq.~(\ref{eq:axialparam})]
and all have different masses.

As $\epsilon^2$ increases, we expect, based on the results of the
previous two sections, that we will enter a phase
which is connected to the CP-violating (pink) phases found above. 
This should have a condensate having
components in both $n_1$ and $n_3$ directions,
and $\theta$ taking non-extremal values.
Indeed, if $\sin\theta$ and $\sin\varphi_1$ are both non-zero,
Eq.~(\ref{eq:partialVtheta1}) is solved by
\begin{equation}
\sin{\theta}\cos\varphi_1 =\frac{\muhat}{2c_\ell \epsilon^2}\,.
\label{eq:arbtwist1}
\end{equation}
This requires that $c_\ell\epsilon^2\ge|\muhat|$.
Inserting this in Eq.~(\ref{eq:partialVthetab}) then yields
\begin{equation}
\cos{\theta}=\frac{\mhat}{2(c_\ell \epsilon^2-w')}\,,
\label{eq:arbtwist2}
\end{equation}
which is valid if $2(c_\ell\epsilon^2-w')\le\mhat$.
The solution given by Eqs.~(\ref{eq:arbtwist1}) and (\ref{eq:arbtwist2})
turns out to give the absolute minimum of the
potential where it is valid. Its boundary with the continuum-like
phase occurs when $|\cos\varphi_1|=1$, and is thus described by
\begin{equation}
\left(\frac{\mhat}{2(c_\ell\epsilon^2-w')}\right)^2
+
\left(\frac{\muhat}{2c_\ell\epsilon^2}\right)^2 = 1
\,.
\end{equation}
For fixed $\epsilon$, this is an ellipse in the $\mhat$, $\muhat$ plane. 
One pion ($\pi_3$) is massless along this critical surface.

Within the CP-violating phase all pions are massive,
with the mass eigenstates being $\pi_2$ and a mixture of $\pi_1$ and $\pi_3$.
The general expressions for these masses are uninformative,
and we quote only the results along the boundary of this phase.
Here, in addition to the massless $\pi_3$ we find 
\begin{align}
&m_{{\pi}_1}^2= 2c_\ell\epsilon^2-
\frac{2w'\muhat^2}{(2c_\ell \epsilon^2)^2} && 
m_{\pi_2}^2= 2c_\ell\epsilon^2\,. 
\end{align}

\bigskip
The only other critical lines are those we found at maximal twist,
namely at $\mhat=0$, $\muhat=2w'$ and $c_\ell\epsilon^2\le w'$.
\bigskip

\begin{figure}[H]
\centering
 \begin{subfigure}{0.49\textwidth}
\includegraphics[scale=.6]{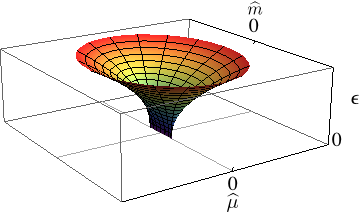}
\end{subfigure}
 \begin{subfigure}{0.49\textwidth}
\includegraphics[scale=.5]{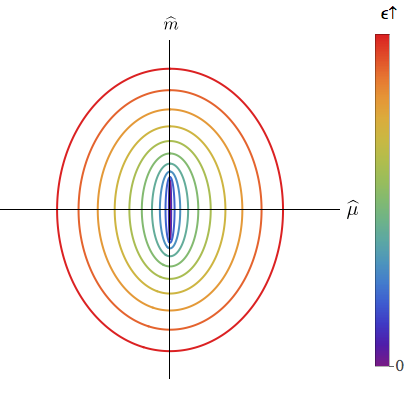}
\caption{Aoki scenario ($w'<0$)} 
\end{subfigure}
 \begin{subfigure}{0.49\textwidth}
\includegraphics[scale=.6]{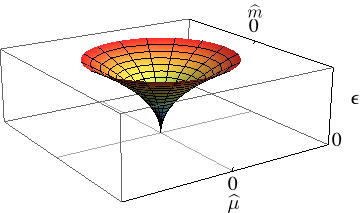}
\end{subfigure}
 \begin{subfigure}{0.49\textwidth}
\includegraphics[scale=.5]{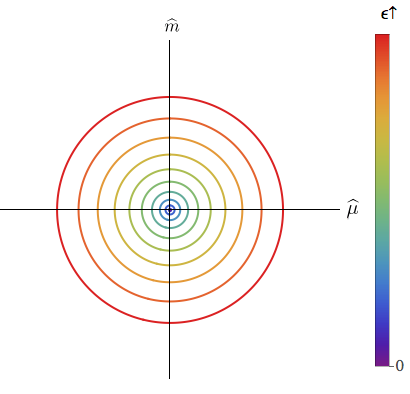}
\caption{Continuum ($w'=0$)} 
\end{subfigure}
 \begin{subfigure}{0.49\textwidth}
\includegraphics[scale=.6]{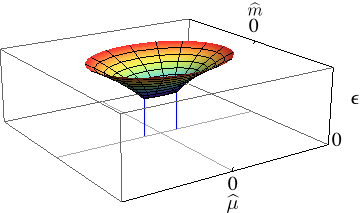}
\end{subfigure}
 \begin{subfigure}{0.49\textwidth}
\includegraphics[scale=.5]{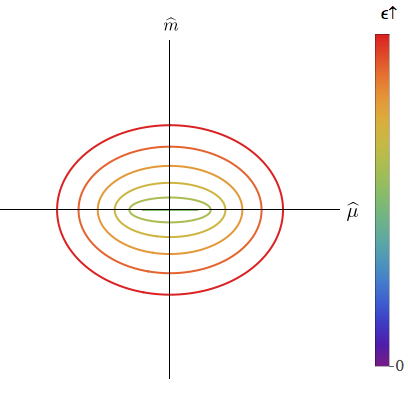}
\caption{First-order scenario ($w'>0$)} 
\end{subfigure}
\caption{\label{fig:generalTwistPhases} 
Location of the critical manifold for arbitrary twist. 
Results are shown only for $\epsilon>0$ since the phase diagrams 
are symmetric under reflection in the $\epsilon=0$ plane.
The left panels show 3-d plots, the right panels contour plots.
For $w'>0$, the contour plots do not include the two critical lines 
which reach down to the $\epsilon=0$ plane.
The scale used for $\mhat$ and $\muhat$ is the same, while that
for $\epsilon$ is arbitrary.
See text for the equations describing the critical manifold.}
\end{figure}

The position of the critical manifold resulting from these considerations
is shown in Fig.~\ref{fig:generalTwistPhases} for both scenarios and
in the continuum. The CP-violating phases lie within the (distorted) 
cone-shaped regions. The contour plots show how the circular contours
of the continuum are distorted by discretization effects into ellipses.
We note that, in the first-order scenario shown in
Fig.~\ref{fig:generalTwistPhases}c,
if one passes through any point
in the rectangular region in the $(\mhat,\epsilon)$
plane between the two critical lines there is a first-order
transition at which the condensate changes discontinuously.

\vfill
\newpage

\section{Higher order}
\label{sec:higher}

In this section we consider the effect on the previous results
of the inclusion of the next highest order terms in our power counting,
i.e. those scaling as $a^3\sim m a$. At this order we can still determine
the vacuum using the classical potential of the chiral theory. 
The ${\cal O}(m a)$ term in this potential is 
standard, see, e.g. Ref.~\cite{Rupak:2002sm}.
The ${\cal O}(a^3)$ terms 
have been discussed for $\epsilon=0$ in Ref.~\cite{Sharpe:2005rq};
the results carry over unchanged to $\epsilon\ne0$
since the first additional term
involving $\epsilon$ scales as $a\epsilon^2$ and is of higher order
in our power-counting.
The relevant additional terms entering the potential are
\begin{equation}
\mathcal{V}_{a^3}= 
-\frac{w f^2}{32 W_0 a}
\tr(\chi^\dagger \Sigma + \Sigma^\dagger \chi)
\tr(A^\dagger \Sigma + \Sigma^\dagger A)
- \frac{w_3 f^2}{(8 W_0 a)^3}
\left[\tr(A^\dagger \Sigma +\Sigma^\dagger A)\right]^3
\,,
\end{equation}
where $w$ and $w_3$ are new LECs.
There is also a term proportional to
$\tr(A^\dagger A)\tr(A^\dagger \Sigma +\Sigma^\dagger A)$,
but this can removed by (yet another) redefinition of $\chi$. 
Inserting our standard parametrization
$\left<\Sigma\right>= \exp(i\theta\vec{n} \cdot \vec{\tau})$,
and combining the results with that from the LO potential,
we obtain
\begin{equation}
-\frac{\mathcal{V}_{a^2,\ell_7,a^3}}{f^2} = 
(\mhat\cos{\theta}+\muhat n_1 \sin{\theta})(1+w \cos{\theta}) 
+c_\ell \epsilon^2 n_3^2 \sin^2{\theta} +w'\cos^2{\theta} +
w_3 \cos^3{\theta}
\,.
\end{equation}
The new LECs should satisfy $|w|\ll 1$ and 
$|w_3|\ll |w'|,|c_\ell\epsilon^2|$
in order to be consistent with our power counting.

We begin by considering the untwisted theory, $\muhat=0$,
where the phase diagram and pion masses can be determined
analytically. In this case $\mhat=\chi_\ell$.
As previously, the potential is minimized with $n_3=1$, so that
\begin{equation}
-\frac{\mathcal{V}_{a^2,\ell_7,a^3}}{f^2} \longrightarrow
\chi_\ell\cos{\theta}(1+w\cos{\theta}) 
+c_\ell \epsilon^2 \sin^2{\theta} +w'\cos^2{\theta} +w_3\cos^3{\theta}
\,.
\end{equation}
The stationary points satisfy
\begin{equation}
\sin\theta\left[
\chi_\ell -2(\chi_\ell w - c_\ell \epsilon^2 + w')\cos{\theta} 
+3w_3 \cos^2{\theta}\right]=0\,,
\label{eq:higherorderstat}
\end{equation}
which is solved by $\sin\theta=0$ 
(i.e. giving the usual continuum solutions with $\cos\theta=\pm 1$)
and by the solutions to the quadratic function of $\cos\theta$ in
parentheses. The latter will lead to CP-violating vacua.

\begin{figure}[tb!]
\centering
\begin{subfigure}{0.49\textwidth}
\includegraphics[scale=.35]{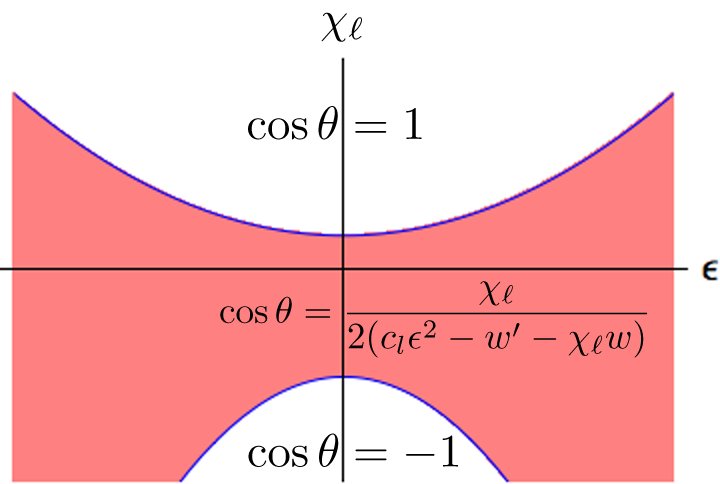}
\caption{Aoki Scenario ($w'<0$)}
\end{subfigure}
\begin{subfigure}{0.49\textwidth}
\includegraphics[scale=.35]{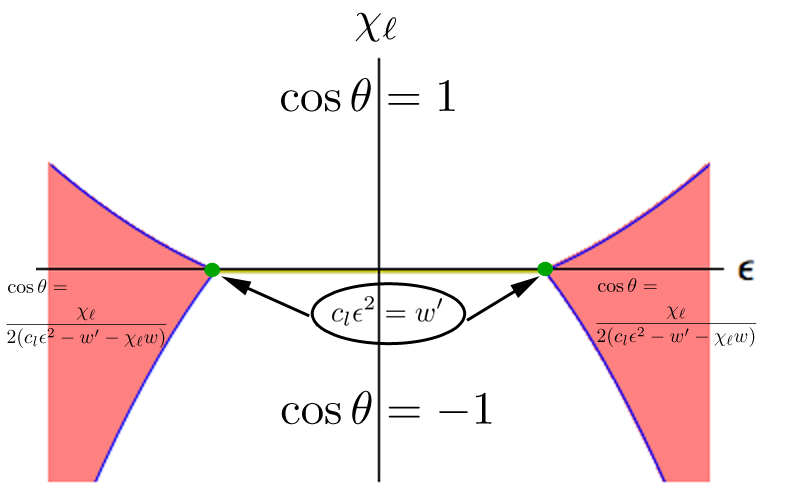}
\caption{First-Order Scenario ($w'>0$)}
\end{subfigure}
\caption{\label{fig:oa3wphases} 
Phase diagrams for untwisted Wilson quarks including the NLO 
${\cal O}(ma)$ term proportional to $w$. 
Compare to LO results in Fig.~\ref{fig:untwistPhases}.}
\end{figure}

To simplify the discussion we consider the impact of the new terms separately.
We first set $w_3=0$. Then we can take $w>0$ without loss of generality,
since simultaneously changing $w\to -w$, $\theta\to\theta+\pi$
and $\chi_\ell\to -\chi_\ell$ leaves the potential unaffected.
As the $w$ contribution to Eq.~(\ref{eq:higherorderstat}) leaves
the function in parentheses linear in $\cos\theta$,
the analysis is little changed from that at LO (see Sec.~\ref{sec:disc}).
We find that the CP-violating solution,
\begin{equation}
\cos{\theta}=\frac{\chi_\ell}{2(c_\ell \epsilon^2 - w'-\chi_\ell w)}
\,,
\end{equation}
minimizes the potential where it is valid, i.e. 
wherever $|\cos\theta|< 1$.
The endpoints of this phase give second-order transitions occurring at masses
\begin{equation}
\chi_\ell  = 
\pm \frac{2(c_\ell \epsilon^2 - w')}{1\pm 2 w}\,.
\end{equation}
Thus the phase boundaries are no longer symmetric with respect to
$\chi_\ell=0$.
As in the LO case, if $w'>0$ and $c_\ell \epsilon^2 < w'$, the
transition becomes first order (with the $w$ term having no impact
since the transition occurs at $\chi_\ell=0$).
The resultant phase diagrams are shown in Fig.~\ref{fig:oa3wphases}.

\begin{figure}[tb!]
\centering
\begin{subfigure}{0.49\textwidth}
\includegraphics[scale=.32]{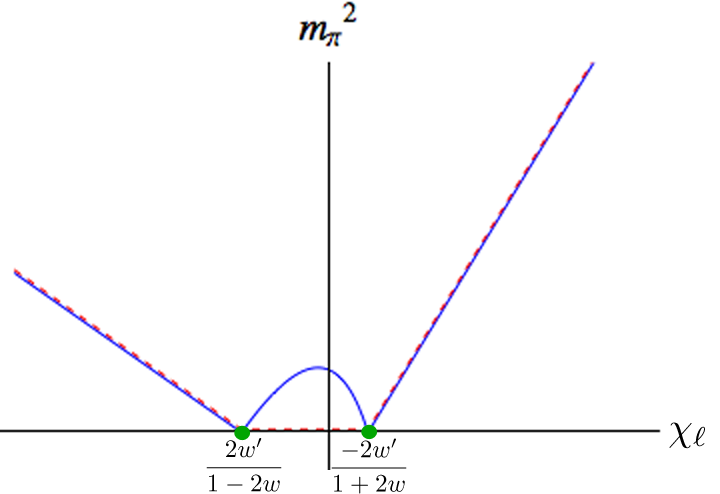}
\caption{$w'<0$, $c_\ell \epsilon^2=0$,  $w>0$}
\end{subfigure}
\begin{subfigure}{0.49\textwidth}
\includegraphics[scale=.32]{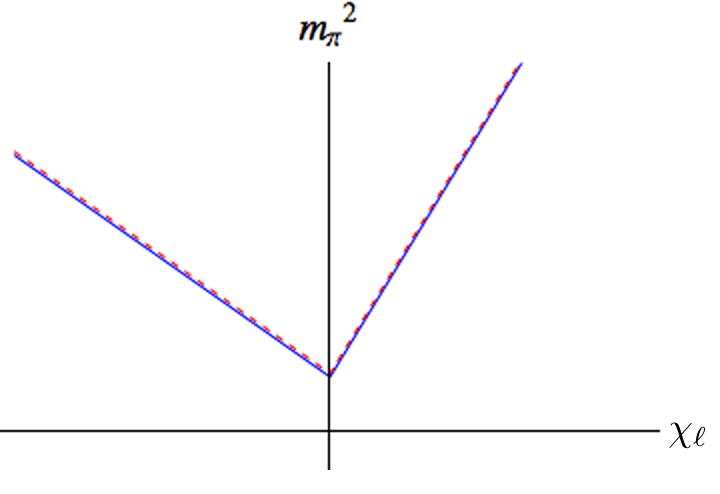}
\caption{$w'>0$, $c_\ell \epsilon^2=0$, $w>0$}
\end{subfigure}
\begin{subfigure}{0.49\textwidth}
\includegraphics[scale=.32]{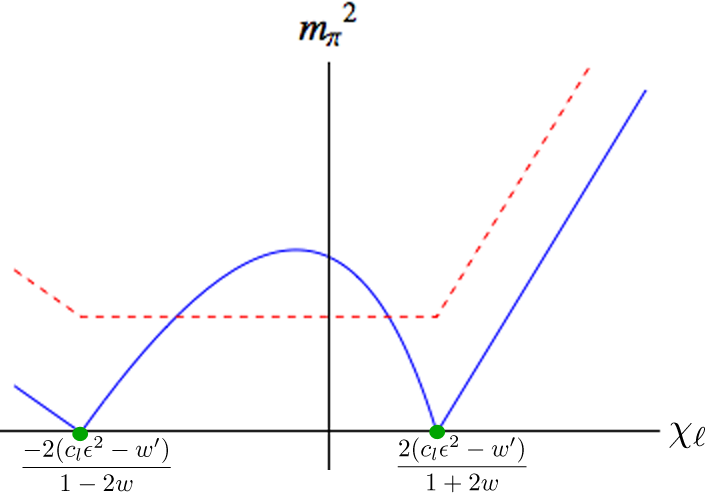}
\caption{$w'<0$, $c_\ell \epsilon^2>0$, $w>0$}
\end{subfigure}
\begin{subfigure}{0.49\textwidth}
\includegraphics[scale=.32]{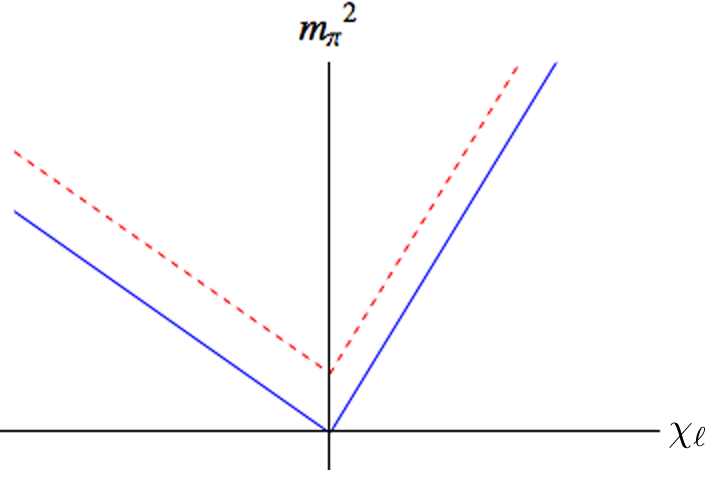}
\caption{ $c_\ell \epsilon^2=w'>0$, $w>0$}
\end{subfigure}
\begin{subfigure}{0.49\textwidth}
\includegraphics[scale=.32]{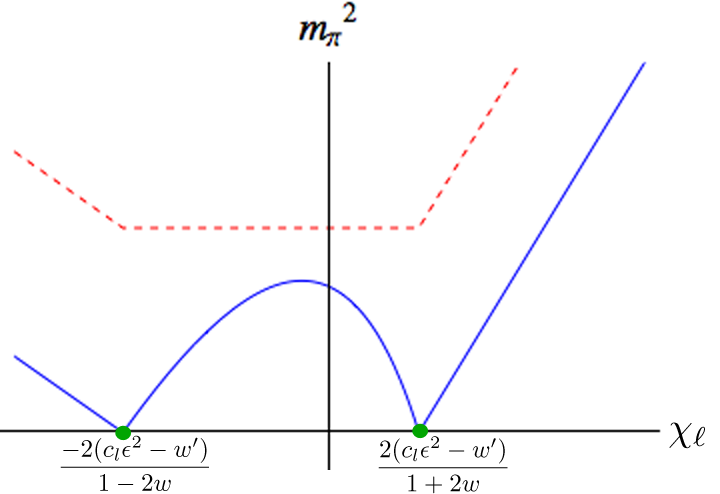}
\caption{ $c_\ell \epsilon^2>w'>0$, $w>0$}
\end{subfigure}
\caption{\label{fig:UnTwistOa3wPiMasses} 
Pion masses for untwisted Wilson fermions including the
effects of the NLO $w$ term with $w>0$ (but with $w_3=0$). 
The figures should be compared to the LO results
in Figs.~\ref{fig:UnTwistPiMasses}(c-g), respectively.
See Fig.~\ref{fig:UnTwistPiMasses} also for notation.}
\end{figure}

We have also calculated the pion masses. In the CP-conserving phases the
results are
\begin{align}
m_{\pi^0}^2&= |\chi_\ell|(1 + {\rm sign}(\chi_\ell) 2w) 
- 2 (c_\ell \epsilon^2-w')\,,
\\
m_{\pi^\pm}^2&= m_{\pi^0}^2 + 2 c_\ell \epsilon^2\,.
\end{align}
The only change from the LO results, Eqs.~(\ref{eq:mpi0untwistLO})
and (\ref{eq:mpipuntwistLO}), is that the slope with respect to
$\chi_\ell$ is no longer symmetric when $\chi_\ell$ changes sign.
In the CP-violating phases we find 
\begin{equation}
m_{\pi^0}^2=
2( c_\ell \epsilon ^2  -w'-\chi_\ell w) \sin^2\theta
\ \ {\rm and}\ \ 
m_{\pi^\pm}^2= 2c_\ell \epsilon^2\,,
\end{equation}
where again only the former result is changed.
The resulting pion masses are shown in Fig.~\ref{fig:UnTwistOa3wPiMasses},
and show clearly the above-mentioned asymmetry.

We now consider the impact of the $w_3$ term, setting $w=0$.
Again, without loss of generality, we can assume $w_3>0$.
The CP-violating stationary points are now obtained 
from Eq.~(\ref{eq:higherorderstat})
by solving a quadratic equation, leading to the solutions
\begin{equation}
\cos{\theta_\pm}= \frac{(c_\ell \epsilon^2 - w') \pm 
\sqrt{(c_\ell \epsilon^2 - w')^2 - 3\chi_\ell w_3}}{3w_3}
\,.
\label{eq:twoCPsol}
\end{equation}
It is straightforward to see from the properties of
a cubic that, since $w_3>0$, only the $\theta_-$ solution
can lead to a minimum of the potential.
Whether it does lead to a minimum
is a more subtle question than in the LO analysis.

We begin by discussing the limit of small $|w_3|$.
Specifically, if we assume
$|c_\ell \epsilon^2-w'|\sim |\chi_\ell| \gg |w_3|$,
the square root in Eq.~(\ref{eq:twoCPsol})
can be expanded in powers of $w_3$.
It is then straightforward to show that one recovers
the LO results aside from small corrections proportional to
$|w_3/(c_\ell \epsilon^2-w')|$.
In particular, if $c_\ell \epsilon^2-w' > 0$ there is
a CP-violating phase ending in second-order transitions  to
continuum-like phases, while if
$c_\ell \epsilon^2-w' < 0$ there is a first-order transition.

The positions of these transitions are, however, shifted
slightly by the $w_3$ term. The boundaries of the CP-violating
phase occur when $\cos\theta_-=\pm 1$ which gives
\begin{equation}
\chi_\ell= \pm 2 (c_\ell \epsilon^2-w') - 3 w_3\,,
\end{equation}
without any ${\cal O}(w_3^2)$ corrections.
In words, the boundaries are simply offset 
from the LO result, Eq.~(\ref{eq:secondboundary}), by $-3 w_3$.
In the first-order scenario, the transition occurs at the
value of $\chi_\ell$ such that the potentials at $\cos\theta=\pm1$ agree.
This happens when
\begin{equation}
\chi_\ell=-w_3\,,
\end{equation}
so that the first-order transition line is offset from
the LO result $\chi_\ell=0$ by $-w_3$
(again, without any higher-order corrections).

More interesting changes occur when 
$|c_\ell\epsilon^2-w'|\sim |w_3|$.
Note that this does not require that $w_3$ be large, but rather
that there is a cancellation between the $c_\ell\epsilon^2$ and
$w'$ terms. 
Here we encounter a phenomenon first noted at $\epsilon=0$
in Ref.~\cite{Sharpe:2005rq}: one can have a {\em first-order}
transition from the continuum-like phase into the CP-violating
phase, followed by a second-order transition to the other
continuum-like phase. This occurs when the local minimum
at $\theta_-$ (with $|\cos\theta_-|<1$ and $\cos\theta_-$ real)
has the same potential as that at $\cos\theta=1$. Then, as
$\chi_\ell$ is reduced, $\theta$ jumps from $\theta=0$ to
$\theta_-$. This is possible with a cubic potential, but not
with a quadratic. Solving
\begin{equation}
\mathcal{V}_{a^2,\ell_7,a^3}(\theta_-)=
\mathcal{V}_{a^2,\ell_7,a^3}(0)
\end{equation}
leads to the following equation for the first-order boundary
\begin{equation}
\chi_\ell = \frac{(w'-c_\ell \epsilon^2-3w_3)(w'-c_\ell\epsilon^2+w_3)}
{4 w_3}\,.
\label{eq:newfirstboundary}
\end{equation}
As one moves along this boundary $\cos\theta_-$ varies.
The boundary ends when either $\cos\theta_-=1$,
so there is no jump in $\theta$, and the transition
becomes second-order, or when $\cos\theta_-=-1$, so there is only a
first-order transition without the subsequent CP-violating phase.
Combining Eqs.~(\ref{eq:twoCPsol}) and (\ref{eq:newfirstboundary})
we find that the transition becomes second-order at
\begin{equation}
\chi_\ell = c_\ell \epsilon^2-w'= 3 w_3\,,
\end{equation}
while it becomes first-order at
\begin{equation}
\chi_\ell = c_\ell \epsilon^2-w'= -w_3\,.
\end{equation}
The first of these equations can be satisfied if $w'> -3 w_3$, and
so reaches from the first-order scenario ($w'>0$) into a
small region of the Aoki scenario.
The second requires $w'> w_3$ and thus occurs only in the first-order
scenario.

\begin{figure}[tb!]
\centering
\begin{subfigure}{1.0\textwidth}
\centering
\includegraphics[scale=.35]{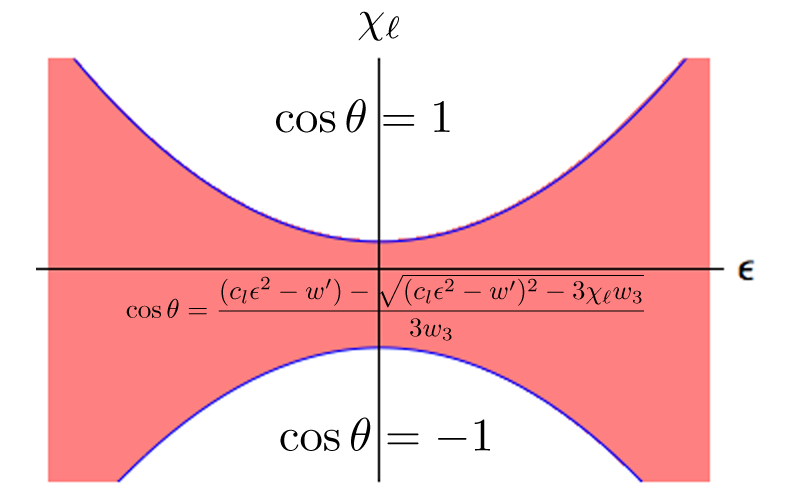}
\caption{\centering Aoki scenario with $w' < -3 w_3 < 0$}
\end{subfigure}
\begin{subfigure}{1.0\textwidth}
\centering
\includegraphics[scale=.35]{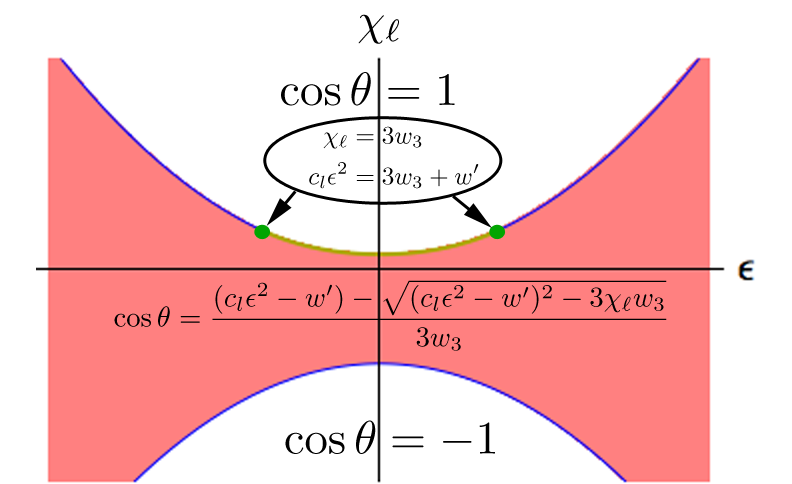}
\caption{\centering Aoki or first-order scenario with $-3 w_3 < w' < w_3$
(and $w_3>0$)}
\end{subfigure}
\begin{subfigure}{1.0\textwidth}
\centering
\includegraphics[scale=.35]{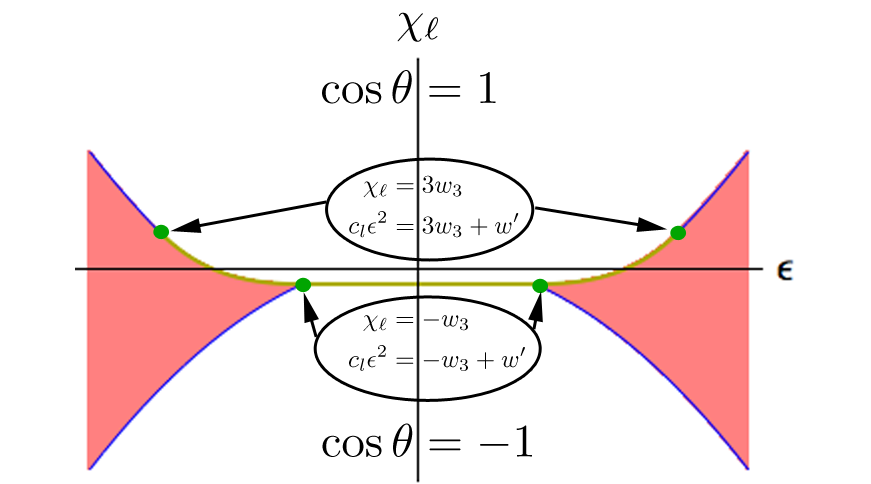}
\caption{\centering First-order scenario with $w'>w_3>0$; $\cos\theta$
 in pink region is as is in (a) and (b)}
\end{subfigure}
\caption{\label{fig:untwistnumericalOa3Phases1} 
Phase diagrams for untwisted Wilson fermions including the NLO 
${\cal O}(a^3)$ term proportional to $w_3$. 
Compare to LO results in Fig.~\ref{fig:untwistPhases}.}
\end{figure}

These results lead to the phase diagrams shown in 
Fig.~\ref{fig:untwistnumericalOa3Phases1}. 
We see that the changes due to the $w_3$ term are more
substantive than those due to the $w$ term.

\begin{figure}[tb!]
\centering
\begin{subfigure}{0.49\textwidth}
\includegraphics[scale=.32]{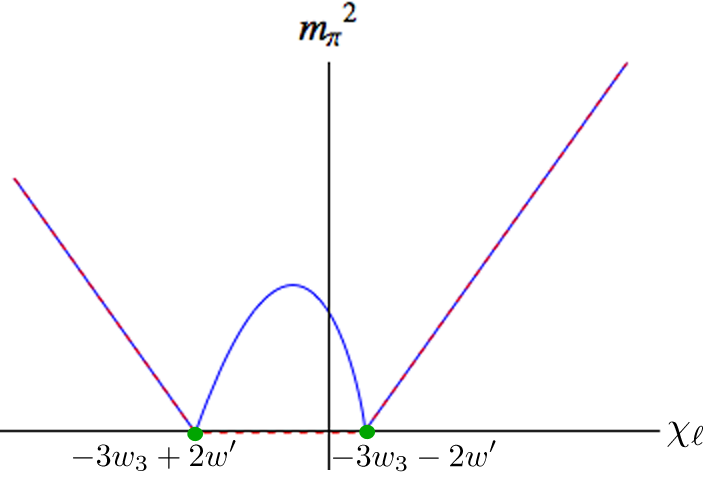}
\caption{ $c_\ell \epsilon^2=0$, $-w'<-3w_3<0$}
\end{subfigure}
\begin{subfigure}{0.49\textwidth}
\includegraphics[scale=.32]{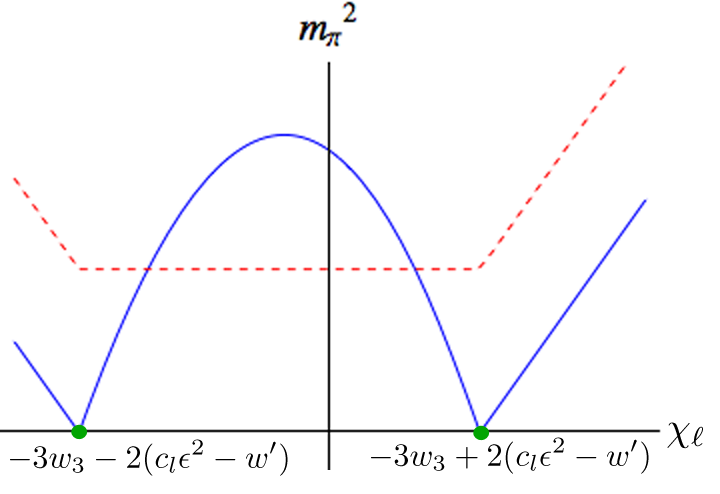}
\caption{ $c_\ell \epsilon^2>0$, $-w'<-3w_3<0$}
\end{subfigure}
\caption{\label{fig:UnTwistOa3w3PiMassesA} 
NLO pion masses for untwisted Wilson fermions with $w_3>0$ and $w=0$.
Results are for the Aoki scenario with $w' < -3 w_3 < 0$,
corresponding to the phase diagram of 
Fig.~\ref{fig:untwistnumericalOa3Phases1}a.
Notation as in Fig.~\ref{fig:UnTwistPiMasses}.}
\end{figure}

\begin{figure}[tb!]
\centering
\begin{subfigure}{0.49\textwidth}
\includegraphics[scale=.32]{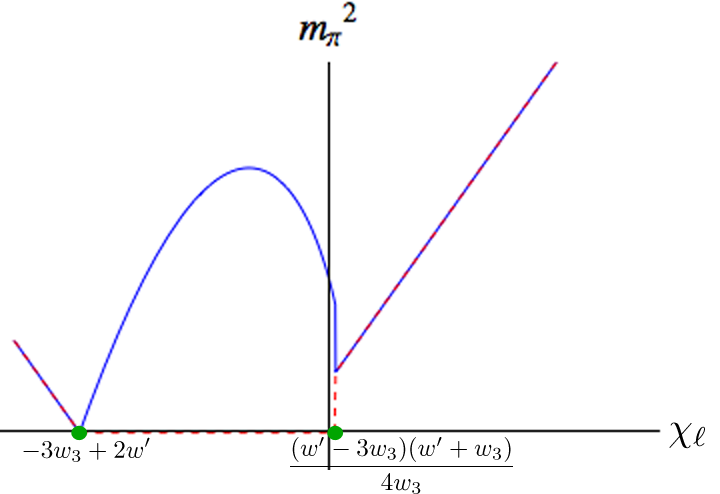}
\caption{$c_\ell \epsilon^2=0$, $-3 w_3 < w' < w_3$}
\end{subfigure}
%
%
\begin{subfigure}{0.49\textwidth}
\includegraphics[scale=.32]{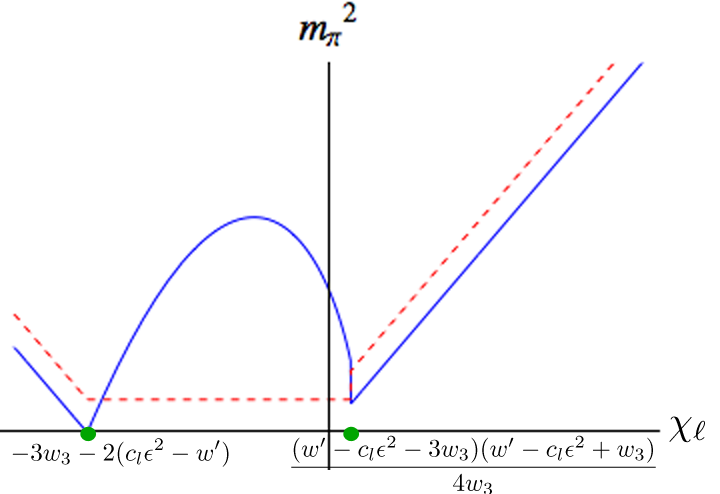}
\caption{$c_\ell \epsilon^2\ge w'+3w_3$, $-3 w_3 < w' < w_3$}
\end{subfigure}
\caption{\label{fig:UnTwistOa3w3PiMassesB} 
Examples of NLO pion masses for untwisted Wilson fermions 
with $w_3>0$ and $w=0$.
Results are for $-3 w_3 < w' < w_3$, corresponding to the phase diagram of 
Fig.~\ref{fig:untwistnumericalOa3Phases1}b.}
\end{figure}

\begin{figure}[tb!]
\centering
\begin{subfigure}{0.49\textwidth}
\includegraphics[scale=.32]{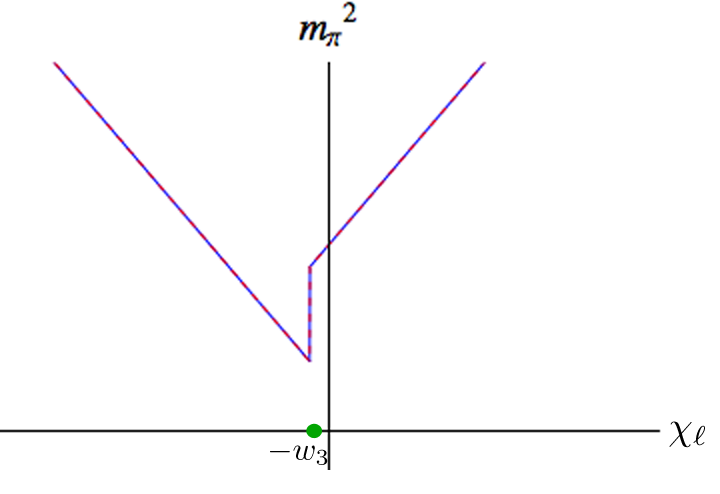}
\caption{ $c_\ell \epsilon^2=0$, $w'> w_3>0$}
\end{subfigure}
\begin{subfigure}{0.49\textwidth}
\includegraphics[scale=.32]{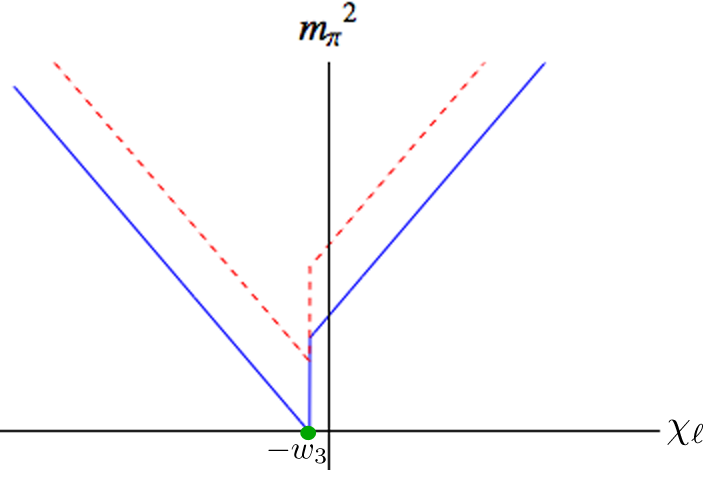}
\caption{ $c_\ell \epsilon^2=w'-w_3$, $w'>w_3>0$}
\end{subfigure}

\begin{subfigure}{0.49\textwidth}
\includegraphics[scale=.32]{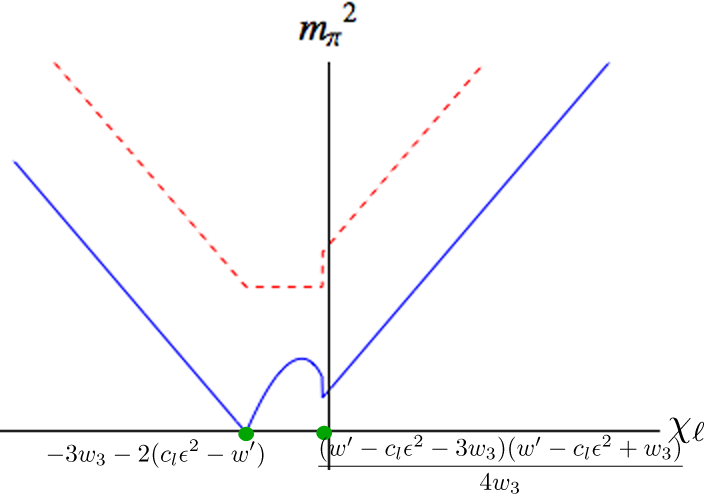}
\caption{ $w'-w_3<c_\ell \epsilon^2< w'+3w_3$, $w'>w_3>0$}
\end{subfigure}
\begin{subfigure}{0.49\textwidth}
\includegraphics[scale=.32]{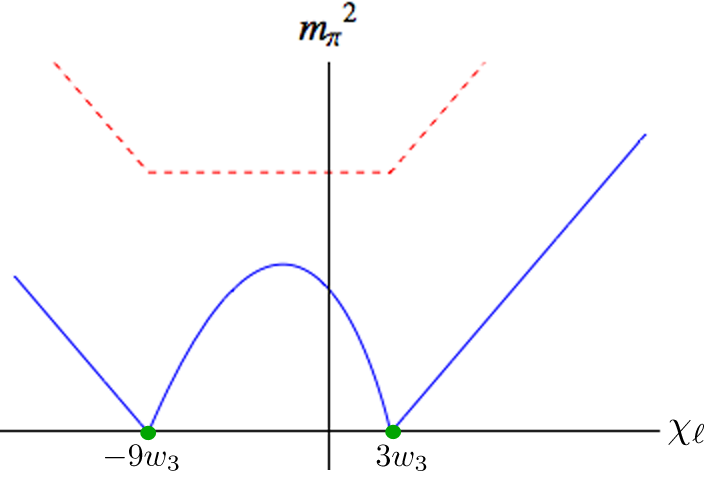}
\caption{ $c_\ell \epsilon^2=w'+3w_3$, $w'>w_3>0$}
\end{subfigure}

\caption{\label{fig:UnTwistOa3w3PiMassesC} 
NLO pion masses for untwisted Wilson fermions with $w_3>0$ and $w=0$.
Results are for the first-order scenario
with $w' > w_3$, corresponding to the phase diagram of 
Fig.~\ref{fig:untwistnumericalOa3Phases1}c.}
\end{figure}

We show the corresponding pion masses in 
Figs.~\ref{fig:UnTwistOa3w3PiMassesA}-\ref{fig:UnTwistOa3w3PiMassesC};
for the sake of brevity we do not quote the analytic forms.
Fig.~\ref{fig:UnTwistOa3w3PiMassesA} shows two ``slices''
through the phase diagram of Fig.~\ref{fig:untwistnumericalOa3Phases1}a.
These should be compared to the LO results in 
Figs.~\ref{fig:untwistPhases}c and \ref{fig:untwistPhases}e, respectively.

In Fig.~\ref{fig:UnTwistOa3w3PiMassesB} we show two slices
through the phase diagram of Fig.~\ref{fig:untwistnumericalOa3Phases1}b.
The first, at $\epsilon=0$, shows the first-order transition,
at which all pion masses are discontinuous.
The charged pions become massless in the CP-violating/Aoki phase,
while the neutral pion is massive.
In the second slice, for which $\epsilon$ satisfies 
$0<c_\ell \epsilon^2< w'+3w_3$, 
the discontinuities remain,
but all pions are massive in the CP-violating phase
(except at the lower boundary where the neutral pion mass vanishes).
Once $c_\ell\epsilon^2\ge w'+3 w_3$, the pion masses behave as
in Fig.~\ref{fig:UnTwistOa3w3PiMassesA}b.

In Fig.~\ref{fig:UnTwistOa3w3PiMassesC} we show four slices
through the phase diagram of Fig.~\ref{fig:untwistnumericalOa3Phases1}c.
The first (at $\epsilon=0$)  shows how the $w_3$ term
leads to a discontinuity in the pion masses at the
first-order transition, unlike at LO.
This was first observed in Ref.~\cite{Sharpe:2005rq}.
For non-zero $\epsilon$, the charged and neutral pions are no longer
degenerate, and both have a discontinuity.
When one reaches $c_\ell \epsilon^2=w'-w_3$, the neutral pion
becomes massless at the transition point, as shown in
the second slice.
This is the beginning of the CP-violating phase. 
As $\epsilon^2$ increases further, one has both first and
second-order transitions, as shown in the third slice.
The final slice shows the value of $\epsilon^2$
at which the first-order transition turns into a second-order transition.
For larger values of $\epsilon^2$ the pion masses behaves as in
Fig.~\ref{fig:UnTwistOa3w3PiMassesA}b.

\begin{figure}[tb!]
\centering

\begin{subfigure}{0.49\textwidth}
\includegraphics[scale=.5]{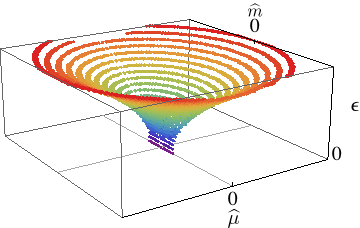}
\end{subfigure}
\begin{subfigure}{0.49\textwidth}
\includegraphics[scale=.5]{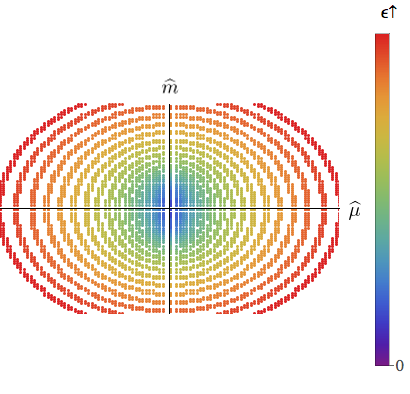}
\caption{$w_3=w=0$}
\end{subfigure}
\begin{subfigure}{0.49\textwidth}
\includegraphics[scale=.5]{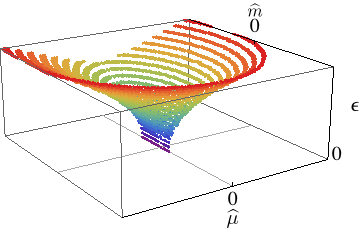}
\end{subfigure}
\begin{subfigure}{0.49\textwidth}
\includegraphics[scale=.5]{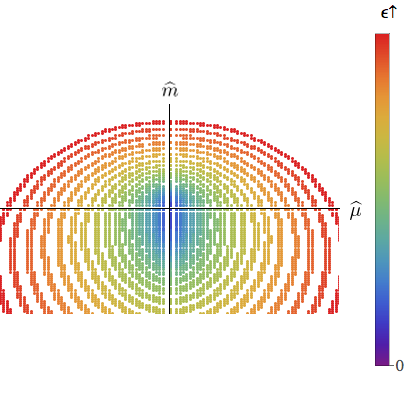}
\caption{$w_3=0$, $w>0$}
\end{subfigure}
\begin{subfigure}{0.49\textwidth}
\includegraphics[scale=.5]{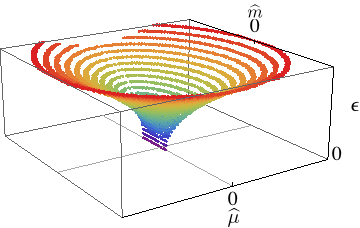}
\end{subfigure}
\begin{subfigure}{0.49\textwidth}
\includegraphics[scale=.5]{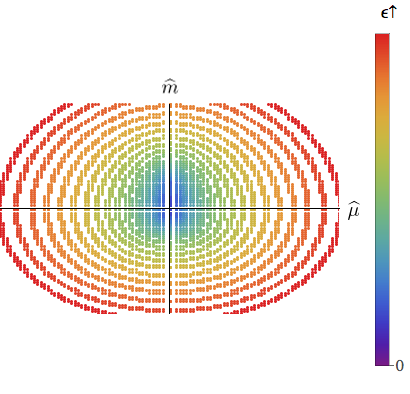}
\caption{$w'< -3w_3<0$, $w=0$}
\end{subfigure}

\caption{\label{fig:numbManifoldAoki}
Location of the critical manifold in the Aoki scenario
($w'<0$) including NLO terms. 
Notation as in Fig.~\ref{fig:generalTwistPhases}.}
\end{figure}

\begin{figure}[tb!]
\centering

\begin{subfigure}{0.49\textwidth}
\includegraphics[scale=.5]{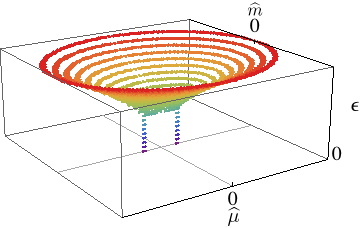}
\end{subfigure}
\begin{subfigure}{0.49\textwidth}
\includegraphics[scale=.5]{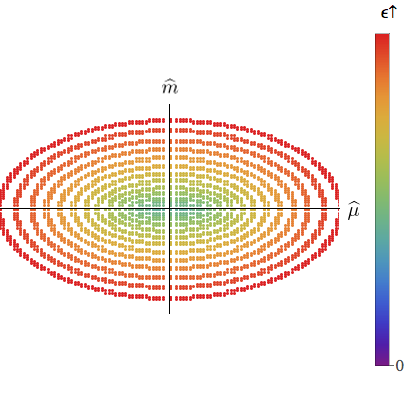}
\caption{$w_3=w=0$}
\end{subfigure}
\begin{subfigure}{0.49\textwidth}
\includegraphics[scale=.5]{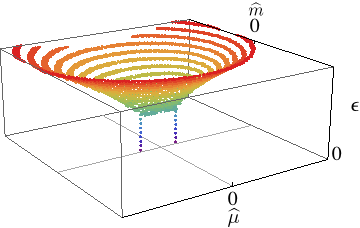}
\end{subfigure}
\begin{subfigure}{0.49\textwidth}
\includegraphics[scale=.5]{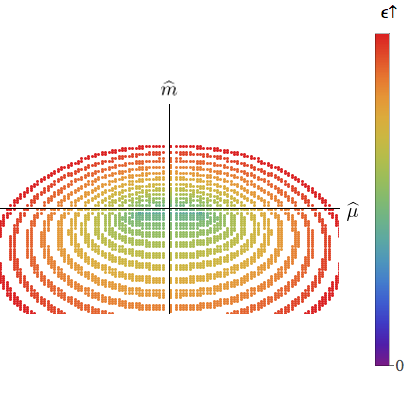}
\caption{$w_3=0$, $w>0$}
\end{subfigure}
\begin{subfigure}{0.49\textwidth}
\includegraphics[scale=.5]{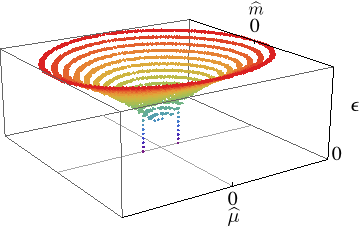}
\end{subfigure}
\begin{subfigure}{0.49\textwidth}
\includegraphics[scale=.5]{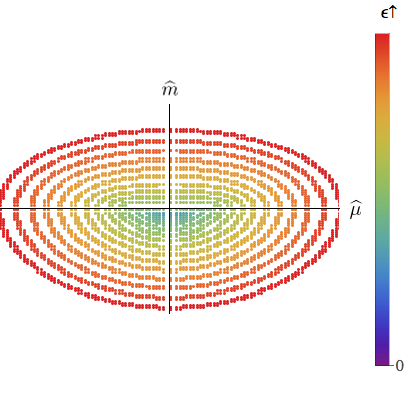}
\caption{$w'>w_3>0$, $w=0$}
\end{subfigure}

\caption{\label{fig:numbManifold}
Location of the critical manifold in the first-order scenario
($w'>0$) including NLO terms. 
Notation as in Fig.~\ref{fig:generalTwistPhases}.}
\end{figure}

The higher-order analysis in the twisted case is more complicated.
Maximal twist no longer occurs, in general, at $\mhat=0$,
so one is forced to do the analysis for both $\mhat$ and $\muhat$
non-vanishing. In practice, this requires numerical minimization of
the potential. The resulting phase diagram and pion masses
for $\epsilon=0$ have been discussed in detail in
Ref.~\cite{Sharpe:2005rq}. The addition of
isospin-breaking leads both to small quantitative changes, 
and to qualitative changes in small regions of the phase plane.
We restrict ourselves here to showing how the NLO
terms impact the critical manifold (on which at least one pion is massless).
The Aoki and first-order scenarios are shown, respectively,
in Figs.~\ref{fig:numbManifoldAoki} and \ref{fig:numbManifold}.

The main effect is a distortion of the elliptical cross sections
of the critical manifold. In addition, the two vertical
critical lines in the first-order scenario are shifted slightly
in position. The most significant qualitative change is the appearance
of a hole in the manifold when $w'>w_3>0$,
which is (barely) visible above the $\muhat=0$ axis in
the right panel of Fig.~\ref{fig:numbManifold}c.
This occurs because of the extended
first-order transition region seen in the untwisted phase diagram 
of Fig.~\ref{fig:untwistnumericalOa3Phases1}c.

We end this section by addressing the
question of whether higher-order effects move 
unphysical phases closer to the point with physical masses.
The answer depends on the sign of $w$ and $w_3$.
For untwisted fermions, the results of 
Figs.~\ref{fig:oa3wphases}-\ref{fig:numbManifold}, show that
positive $w$ and $w_3$ move unphysical phases away from the physical point.
Conversely, negative values of these LECs would move the phases closer.
For twisted-mass fermions the answer is more complicated,
depending on the choice of twist angle.

\clearpage

\section{Conclusions}
\label{sec:concl}

In this work we have studied how using non-degenerate
up and down quarks changes the phase structure caused 
by competition between quark mass and discretization effects.
We draw two main conclusions.
First, the continuum CP-violating phase
is continuously connected to the Aoki phase induced by
discretization effects.
Second, discretization effects can move
the theory with physical quark masses closer to, 
or even into, unphysical phases.
Whether this happens depends both on the twist angle
and on the details of the discretization
(the latter impacting the values of the LECs $w'$, etc.).
Our overall message is that a complicated phase structure
lies in the vicinity of the physical point and simulations 
should be careful to avoid unphysical phases.

For twisted mass fermions our results for pion masses extend
those of Ref.~\cite{Munster:2006yr} into the Aoki regime
($m\sim a^2$).
In the continuum-like phase, with both
twisting and non-degeneracy, the eigenstates are
$\pi_1$, $\pi_2$ and $\pi_3$, with all three
pions having different masses. Our formulae may be of
use in removing the discretization effects from masses
determined in simulations, although we stress again that ${\cal O}(m^2)$
terms dropped in our power counting may be important if
precision fitting is required.

One shortcoming of this work is that it does not include
electromagnetic effects. In the pion sector, these
lead to isospin breaking that is generically larger than that from
quark non-degeneracy, and can also impact the phase structure.\footnote{%
Another generalization that one can consider is the inclusion of an
isospin chemical
potential. This has been discussed recently for degenerate quarks
in Ref.~\cite{Kieberg:lat14}.}
We will discuss the impact of electromagnetism in an upcoming 
work~\cite{inprep},
building upon the recent analysis of Ref.~\cite{Golterman:2014yha}.

\section*{Acknowledgments}
We thank Mario Kieberg and Jac Verbaarschot for discussions and comments.
This work was facilitated through the use of advanced computational,
storage, and networking infrastructure provided by the Hyak
supercomputer system at the University of Washington.
This work was supported in part by the United States Department of Energy 
grants DE-FG02-96ER40956 and DE-SC0011637.

\bibliography{ref}

\end{document}